\documentclass[twocolumn, trackchanges]{aastex631}

\received{February 6, 2023}
\revised{April 28, 2023}
\accepted{April 30, 2023}
\reportnum{Accepted to ApJ}

\shorttitle{Over 500 Days in the Photosphere of SN\,2014dt}
\shortauthors{Camacho-Neves et al.}
\usepackage{longtable}
\usepackage{siunitx}
\usepackage{apjfonts}
\usepackage{enumitem}

\newcommand\R[1]{{\bf #1}} 

\newcommand{\tardis}{\texttt{TARDIS}}
\newcommand{\vouter}{$v_{\rm outer}$}
\newcommand{\vphot}{$v_{\rm phot}$}

\usepackage{soul}

\begin{document}

\title{Over 500 Days in the Life of the Photosphere of the Type Iax Supernova SN\,2014dt}

\correspondingauthor{Yssavo Camacho-Neves}
\email{camacho@physics.rutgers.edu}

\author[0000-0002-9830-3880]{Yssavo Camacho-Neves}
\affiliation{Department of Physics \& Astronomy, Rutgers, the State University of New Jersey, Piscataway, NJ 08854, USA}

\author[0000-0001-8738-6011]{Saurabh W. Jha}
\affiliation{Department of Physics \& Astronomy, Rutgers, the State University of New Jersey, Piscataway, NJ 08854, USA}

\author[0000-0003-4769-4794]{Barnabas Barna}
\affiliation{Department of Experimental Physics, Institute of Physics, University of Szeged, D\'om t\'er 9, Szeged, Hungary}

\author[0000-0002-5995-9692]{Mi Dai}
\affiliation{Department of Physics \& Astronomy, Johns Hopkins University, Baltimore, MD 21218, USA}

\author[0000-0003-3460-0103]{Alexei V. Filippenko}
\affiliation{Department of Astronomy, University of California, Berkeley, CA 94720-3411, USA}

\author[0000-0002-2445-5275]{Ryan J.\ Foley}
\affiliation{Department of Astronomy and Astrophysics, University of California, Santa Cruz, CA 95064, USA}

\author[0000-0002-0832-2974]{Griffin Hosseinzadeh}
\affiliation{Steward Observatory, University of Arizona, 933 North Cherry Avenue, Tucson, AZ 85721-0065, USA}

\author[0000-0003-4253-656X]{D.\ Andrew Howell}
\affiliation{Las Cumbres Observatory, Goleta, CA 93117, USA}
\affiliation{Department of Physics, University of California, Santa Barbara, CA 93106, USA}

\author[0000-0001-5975-290X]{Joel~Johansson}
\affiliation{Oskar Klein Centre, Department of Physics, Stockholm University, AlbaNova, SE-10691 Stockholm, Sweden}

\author[0000-0003-3142-997X]{Patrick L. Kelly}
\affiliation{Minnesota Institute for Astrophysics, University of Minnesota, 116 Church Street SE, Minneapolis, MN 55455, USA}

\author[0000-0002-0479-7235]{Wolfgang E. Kerzendorf}
\affiliation{Department of Physics and Astronomy, Michigan State University, East Lansing, MI 48824, USA}
\affiliation{Department of Computational Mathematics, Science, and Engineering, Michigan State University, East Lansing, MI 48824, USA}

\author[0000-0003-3108-1328]{Lindsey A.\ Kwok}
\affiliation{Department of Physics \& Astronomy, Rutgers, the State University of New Jersey, Piscataway, NJ 08854, USA}

\author[0000-0003-2037-4619]{Conor Larison}
\affiliation{Department of Physics \& Astronomy, Rutgers, the State University of New Jersey, Piscataway, NJ 08854, USA}

\author[0000-0002-0629-8931]{Mark R. Magee}
\affiliation{School of Physics, Trinity College Dublin, University of Dublin, Dublin 2, Ireland}
\affiliation{Institute of Cosmology and Gravitation, University of Portsmouth, Burnaby Road, Portsmouth, PO1 3FX, UK}

\author[0000-0001-5807-7893]{Curtis McCully}
\affiliation{Las Cumbres Observatory, Goleta, CA 93117, USA}
\affiliation{Department of Physics, University of California, Santa Barbara, CA 93106, USA}

\author[0000-0003-3615-9593]{John T. O'Brien}
\affiliation{Department of Physics and Astronomy, Michigan State University, East Lansing, MI 48824, USA}

\author[0000-0001-8415-6720]{Yen-Chen Pan}
\affiliation{Graduate Institute of Astronomy, National Central University, 300 Jhongda Road, Zhongli, Taoyuan, 32001, Taiwan}

\author[0000-0002-2499-9205]{Viraj Pandya}
\affiliation{Center for Computational Astrophysics, Flatiron Institute, New York, NY 10011, USA}
\affiliation{Columbia Astrophysics Laboratory, Columbia University, 550 West 120th Street, New York, NY 10027, USA}
\affiliation{Hubble Fellow}

\author[0000-0002-8310-0829]{Jaladh Singhal}
\affiliation{California Institute of Technology, IPAC, M/C 100-22, Pasadena, CA 91125, USA}

\author[0000-0002-3169-3167]{Benjamin E. Stahl}
\affiliation{Department of Astronomy, University of California, Berkeley, CA 94720-3411, USA}
\affiliation{Department of Physics, University of California, Berkeley, CA 94720-7300, USA}

\author[0000-0003-4610-1117]{Tam\'as Szalai}
\affiliation{Department of Experimental Physics, Institute of Physics, University of Szeged, D\'om t\'er 9, Szeged, Hungary}
\affiliation{ELKH-SZTE Stellar Astrophysics Research Group, H-6500 Baja, Szegedi \'ut Kt. 766, Hungary}

\author{Meredith Wieber}
\affiliation{Minnesota Institute for Astrophysics, University of Minnesota, 116 Church Street SE, Minneapolis, MN 55455, USA}

\author[0000-0003-2544-4516]{Marc Williamson}
\affiliation{Department of Physics, New York University, New York, NY 10003, USA}

\begin{abstract} 
Type Iax supernovae (SN Iax) are the largest known class of peculiar white dwarf supernovae, distinct from normal Type Ia supernovae (SN Ia). The unique properties of SN Iax, especially their strong photospheric lines out to extremely late times, allow us to model their optical spectra and derive physical parameters for the long-lasting photosphere. We present an extensive spectral timeseries, including 21 new spectra, of SN Iax 2014dt from +11 to +562 days after maximum light. We are able to reproduce the entire timeseries with a self-consistent, nearly unaltered deflagration explosion model from \citet{Fink2014MNRAS} using \tardis, an open-source radiative transfer code \citep{Kerzendorf2014MNRAS.440..387K,kerzendorf_wolfgang_2023_7525913}. We find that the photospheric velocity of SN\,2014dt slows its evolution between +64 and +148 days, which closely overlaps the phase when we see SN\,2014dt diverge from the normal spectral evolution of SN~Ia (+90 to +150 days). The photospheric velocity at these epochs, $\sim$400--1000 km s$^{-1}$, may demarcate a boundary within the ejecta below which the physics of SN Iax and normal SN Ia differ. Our results suggest that SN\,2014dt is consistent with a weak deflagration explosion model that leaves behind a bound remnant and drives an optically thick, quasi-steady-state wind creating the photospheric lines at late times. The data also suggest that this wind may weaken at epochs past +450 days, perhaps indicating a radioactive power source that has decayed away.
\end{abstract}


\section{Introduction} \label{sec:intro}

Type Ia supernovae (SN Ia) are important ``standard candles'' in cosmology and a major source of chemical enrichment in the Universe. Despite their importance, a complete understanding of the progenitor system and explosion mechanics of SN Ia remains elusive. The use of SN Ia as standard candles implies that they emerge from a homogeneous origin, but an increasing number of observations show there is diversity within the group. The largest class of peculiar versions of SN Ia -- with over sixty known members -- are Type Iax supernovae \citep[SN Iax;][]{Foley2013ApJ, Jha2017}. With an occurrence of approximately 15--30\% of the SN~Ia rate \citep{Foley2013ApJ,Miller:2017,Srivastav_2022MNRAS.511.2708S}, it is imperative to improve our understanding of this growing SN Iax population.

SN Iax, also called SN 2002cx-like supernovae \citep{Li2003,Filippenko2003_fthp.conf..171F,Jha2006,Phillips2007}, start off resembling normal SN Ia; both have early-time optical spectra dominated by iron-group elements (IGEs) like Cr, Fe, Co, and Ni, as well as intermediate-mass elements (IMEs) like Si, Na, and Ca. The biggest differences at early times are that SN Iax have strongly mixed, low-velocity ejecta and lower luminosities relative to normal SN Ia. Nonetheless, SN Iax are not a homogeneous class. They cover a wide range of peak absolute magnitudes, ranging from nearly as bright as normal SN Ia to objects 100 times fainter \citep{Jha2017}. SN Iax also cover a broad range of photospheric velocities, \vphot, $\sim$ 2000--8000 km s$^{-1}$ \citep{Foley2013ApJ} at maximum light, lower than typical SN Ia velocities, $\sim$ 9000--15,000 km s$^{-1}$ \citep{Silverman2015_10.1093/mnras/stv1011}. 

Despite some early-time similarities, SN Iax spectra diverge significantly from those of normal SN Ia at late times (traditionally defined as $\gtrsim$100 days past maximum light; \citealt{Bowers1997, Branch2008, Silverman2013, Friesen2014, Black2016}). Normal SN Ia are nebular at late times, dominated by forbidden emission lines of Fe, Co, and Ni. SN Iax have not been observed to go fully nebular; though they do exhibit some forbidden lines, their late-time spectra continue to be dominated by permitted features of Fe, Ca, and Na \citep{Jha2006, Sahu2008, Foley2010ApJ, Foley2016}. Not only is this distinct from normal SN~Ia, it is unlike any other class of supernovae.

It is difficult to observe a SN during the full intermediate phase between early and late times ($\sim$ +80 to +200 days) since a SN discovered in the night sky will generally have moved to a region of unobservable daytime sky during this time. Because of sparse observations, the precise epochs when SN Iax spectra transition away from those of normal SN Ia is not well studied. In this paper we analyze SN Iax 2014dt, which was fortuitously found in a very easterly location in the sky and has extensive spectral coverage throughout the intermediate phase. 

SN\,2014dt was discovered at 13.6 mag on 2014 October 29.8 (UT dates are used throughout this paper) in the nearby galaxy M61 \citep[$d \approx 14.6$ Mpc; see \autoref{sec:obs};][]{Nakano2014CBET.4011....1N}. It was classified as a SN Iax by \citet{Ochner2014ATel.6648....1O} using the Asiago Copernico Telescope on 2014 October 31. Although SN\,2014dt was discovered past peak brightness, several attempts have been made to estimate its date of maximum. In this paper, we adopt the $B_{\rm max}$ date from \citet{Kawabata2018PASJ...70..111K}, 2014 October 20.4 (JD = 2,456,950.9 $\pm$ 4.0 days), which is consistent with other measurements \citep{Foley2016,2016ApJ.Fox...816L..13F,Singh2018MNRAS.474.2551S}. 

SN\,2014dt is a moderately luminous SN Iax ($M_B$ between approximately $-$17 and $-$18 at maximum light, depending on the assumed distance), though still dimmer than normal SN Ia with similar decline rates \citep{Kawabata2018PASJ...70..111K, Singh2018MNRAS.474.2551S}. Its spectra resemble those of other SN Iax at both early and late times \citep{Kawabata2018PASJ...70..111K, Singh2018MNRAS.474.2551S}, placing it solidly within the SN Iax class.

Deep pre-explosion images from the \emph{Hubble Space Telescope (HST)} at the location of SN\,2014dt did not detect a progenitor system, which excludes a high-mass main-sequence star ($M_{\text{init}} \gtrsim 16$ M$_{\odot}$) or a high-mass evolved star ($M_{\text{init}} \gtrsim 8$ M$_{\odot}$) as the companion \citep{Foley2015_progenitor_14dt}. We note that another SN Iax, SN\,2012Z, is the only thermonuclear SN with a direct observation of a progenitor system in a pre-explosion image which was interpreted as a luminous, blue, helium-star companion \citep{McCully:2014nature}. This direct detection implies that SN~Iax can arise from a single-degenerate progenitor scenario.

Pure deflagration models -- explosions propagated with only a subsonic flame -- can explain the lower luminosities of SN Iax as well as their unique late-time spectra \citep{Branch2004,Phillips2007,Jha2017}. In particular, a weak deflagration model in which a Chandrasekhar-mass carbon-oxygen white dwarf (WD) explodes but does not completely unbind the star has been especially promising \citep{Jordan2012, Kromer2013MNRAS, Fink2014MNRAS}. In this model, a gravitationally bound remnant -- a most peculiar clump of surviving matter -- is left over, in contrast to normal SN Ia that are thought to fully disrupt their progenitor WD. 
\citet{2016ApJ.Fox...816L..13F} discovered a mid-infrared (MIR) excess in SN\,2014dt at late times, which they interpreted as dust emission. \citet{Foley2016} find a lack of evidence for dust and suggest that the MIR excess may instead come from a bound remnant driving an optically thick wind. 

Other possible explosion models for SN Iax include the core-collapse (CC) model \citep{Valenti2009}, the fallback CC model \citep{Moriya2010ApJ}, and the merger model of a carbon-oxygen and oxygen-neon WD \citep{Kashyap_2018ApJ...869..140K,Karambelkar_2021ApJ...921L...6K}.

We model the optical spectra of SN\,2014dt, from +11 to +562 days after $B_{\rm max}$, using a nearly unaltered deflagration explosion model from \citet{Fink2014MNRAS}. To our knowledge, this is the latest-epoch spectrum modeled for any type of SN using only permitted lines. We describe our observations in \autoref{sec:obs}, and explore the spectral divergence of SN\,2014dt compared to normal SN~Ia in \autoref{sec:diverg}. In \autoref{sec:synthetic_spec} we describe our spectral modeling using the radiative transfer code \tardis\ \citep{Kerzendorf2014MNRAS.440..387K,kerzendorf_wolfgang_2023_7525913}, including analysis of different explosion models and the evolution of the photospheric velocity. We summarize our results and possible interpretations in \autoref{sec:summary}.

\section{Spectral Observations and Reduction}\label{sec:obs}

We present 21 new spectra of SN\,2014dt, ranging from +32 to +456 days past $B_{\rm max}$.

Eleven new spectra of SN\,2014dt were taken using the Robert Stobie Spectrograph \citep[RSS;][]{Smith2006RSS} on the 10~m Southern African Large Telescope (SALT). We used a 1.5\arcsec\ slit (behind an atmospheric dispersion compensator, ADC) and the PG0900 grating, typically tilted to four positions with appropriate order-blocking filters, to cover the entire optical wavelength range without detector chip gaps. The SALT/RSS data were reduced using a custom pipeline (RUSALT) based on PySALT \citep{Crawford2010PySALT}, including wavelength calibration, one-dimensional (1D) extraction, relative flux calibration with spectrophotometric standard stars, combination of the grating tilt positions, telluric absorption removal, and heliocentric correction.

Four new spectra of SN\,2014dt were taken with the DEep Imaging Multi-Object Spectrograph \citep[DEIMOS;][]{DEIMOS2003} on the 10~m Keck II telescope at the W.~M. Keck Observatory. The first of these spectra was taken on 2014-11-21 using a 0.8\arcsec\ slit and the 1200G grating (+GG455 blocking filter) centered at 6170 \AA. These data were reduced with the open-source software PypeIt \citep{pypeit:joss_pub, pypeit:zenodo}. Two DEIMOS spectra (2014-12-20 and 2015-03-18) used the 600ZD grating (+GG455 blocking filter) centered at 7200 \AA, with a 1.2\arcsec\ and 1.0\arcsec\ slit, respectively. One additional DEIMOS spectrum was taken on 2015-12-16 with a 1.0\arcsec\ slit and the 600ZD grating (+GG455 blocking filter) centered at 7000 \AA. These last three spectra were processed with the DEEP2 DEIMOS data-reduction pipeline \citep{2012ascl.soft03003C_DEEP2, 2013_Newman_ApJS..208....5N}.

Four new spectra of SN\,2014dt were taken with FLOYDS, a pair of nearly identical, low resolution spectrographs installed on the 2~m Faulkes Telescope North (FTN) at Haleakal$\bar{\text{a}}$ Observatory and Faulkes Telescope South (FTS) at Siding Spring Observatory; both telescopes are part of the Las Cumbres Observatory Global Telescope Network. The three observations taken on FTN (2015-04-08, 2015-04-19, 2015-05-20) used a 1.6\arcsec\ slit and one observation on FTS (2015-04-24) used a 2.0\arcsec\ slit. All four spectra were observed at the parallactic angle \citep{Filippenko1982PASP...94..715F}, and reduced with the FLOYDS pipeline\footnote{\url{https://github.com/LCOGT/floyds_pipeline}} \citep{Valenti_2014MNRAS.438L.101V}.

One new spectrum taken with the Gemini Multi-Object Spectrograph \citep[GMOS;][]{GMOS2004} on the 8.1~m Gemini-North telescope is a combination of eight exposures from two consecutive nights. All exposures were taken at the parallactic angle, and used the R400 grating with G5305 filter; half at a central wavelength of 6900~\AA\ and half at 7100~\AA. After standard data reduction with \texttt{IRAF}, we applied our own IDL routines to flux calibrate the data and remove telluric lines using the well-exposed continua of spectrophotometric standard stars \citep{Wade_1988ApJ...324..411W,Foley_2003PASP..115.1220F}. Details of the reduction can be found in \citet{Silverman_2012MNRAS.425.1789S}.

One new spectrum was taken with the Goodman High Throughput Spectrograph \citep[GHTS;][]{Goodman_10.1117/12.550069} on the 4~m Southern Astrophysical Research Telescope (SOAR) on 2015-04-10. It used a 1.07\arcsec\ slit, a SYZY 400 grating, and was reduced with the same techniques as other GHTS spectra in \citet{Foley2016}, following \citet{Silverman2012dApJ}. 

All new spectra have minimal differential refraction losses due to being observed at the parallactic angle, using an ADC, and/or being taken at low airmass ($<$ 1.45).

We also collect published optical spectra of SN\,2014dt, ranging from +11 to +562 days past $B_{\rm max}$, from \citet{Ochner2014ATel.6648....1O}, \citet{Foley2015_progenitor_14dt}, \citet{Foley2016}, \citet{Lyman_10.1093/mnras/stx2414}\footnote{J.~Lyman (private communication) provided a host-galaxy-subtracted VLT/MUSE spectrum (2016-01-20) using a background annulus around the source. The latest Keck spectrum (2016-05-07) also had host contamination that we modeled and subtracted to isolate the SN flux, better matching other late-time spectra.}, \citet{Kawabata2018PASJ...70..111K}, \citet{Singh2018MNRAS.474.2551S}\footnote{For two HCT spectra, 2015-02-01 and 2015-02-05, we manually rescaled the red and blue channels to eliminate a flux discontinuity and to better match other spectra at similar epochs.}, and \citet{Stahl_2020MNRAS.492.4325S}. These spectra were obtained, in part, through the Open Supernova Catalog \citep{2017ApJ_opensupernovacatalog...835...64G}, WISeREP\footnote{\url{https://wiserep.weizmann.ac.il}} \citep{2012PASP.Ofer..124..668Y}, and the UC Berkeley Filippenko Group's Supernova Database\footnote{\url{http://heracles.astro.berkeley.edu/sndb}} \citep{Silverman_2012MNRAS.425.1789S}.

\begin{figure*}[htp]
\centering
\includegraphics[width=.97\textwidth]{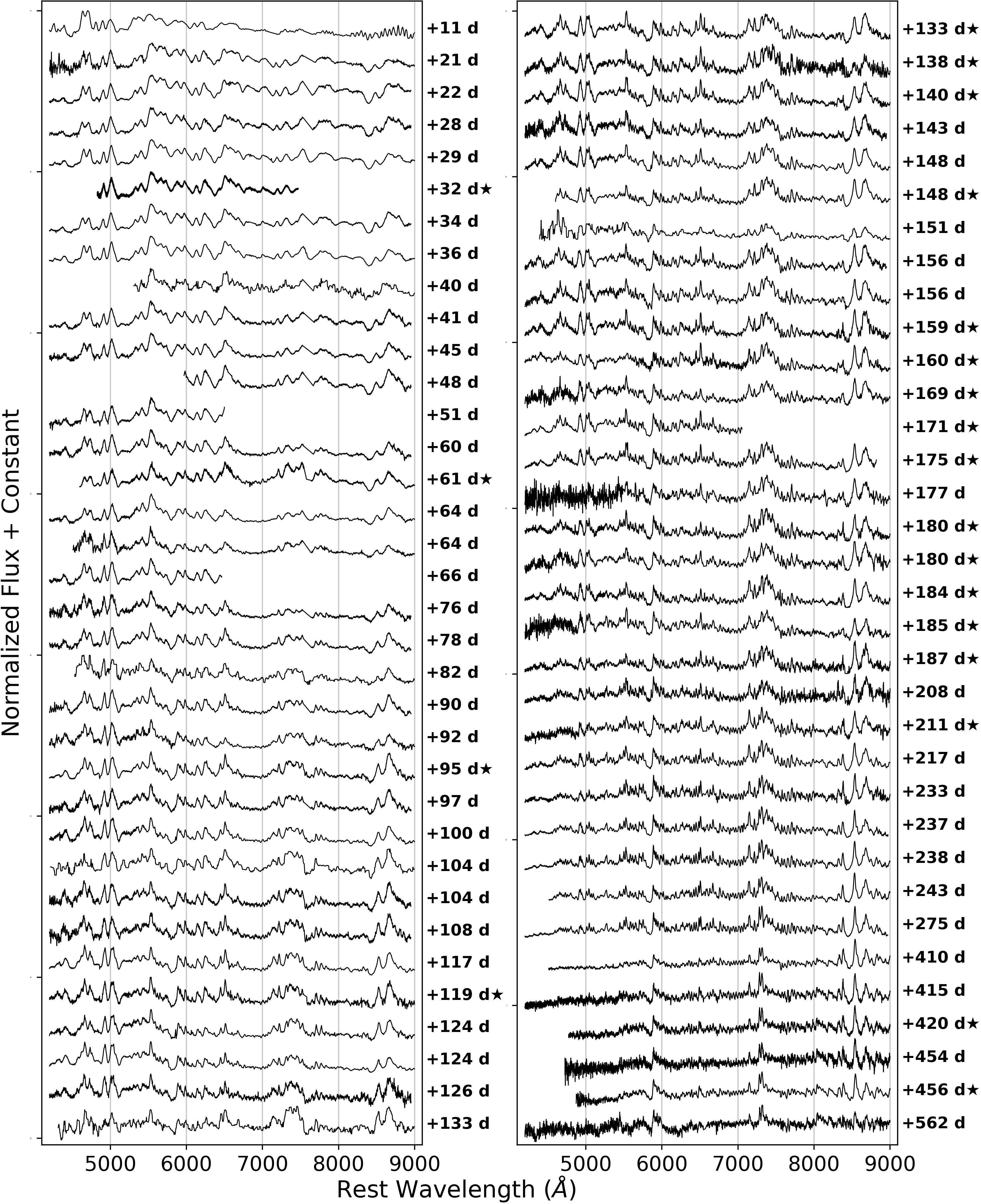}
\caption{Optical spectral evolution of SN\,2014dt. Stars ($\star$) denote new spectra presented in this paper. Phases are calculated relative to the time of $B_{\rm max}$ = 2014 October 20.4  \citep{Kawabata2018PASJ...70..111K} and corrected for time dilation to the host-galaxy rest frame. The spectra are available as data behind the figure.}
\label{fig: all_spec_14dt}
\end{figure*}

\clearpage

\begin{longtable*}{cllccccc}
\caption{Spectral Observations of SN\,2014dt} \label{tab:obs} \\

\hline \hline 
Phase$^{a}$ & UT Date & MJD & Telescope & Exposure & Resolution & Scaled/Mangled & Reference$^{b}$\\ 
 (days) &  &  & /Instrument & (s) &  ($R = \lambda/\Delta \lambda$) & To Filter(s) & \\ \hline 
\endfirsthead

\multicolumn{8}{c}
{{\bfseries \tablename\ \thetable{} -- (continued)}} \\
\hline \hline 
Phase$^{a}$ & UT Date & MJD & Telescope & Exposure & Resolution & Scaled/Mangled & Reference$^{b}$\\ 
 (days) &  &  & /Instrument & (s) &  ($R = \lambda/\Delta \lambda$) & To Filter(s)&  \\ \hline 
\endhead

\hline 
\multicolumn{8}{r}{{Continued on next page}} \\ 
\endfoot

\hline
\multicolumn{8}{l}{{$^{a}$ Days since $B$ maximum, UT 2014 October 20.4 (MJD 56950.4), corrected for time dilation }} \\
\multicolumn{8}{l}{{$^{b}$ 1: \citet{Foley2016}, 2: \citet{Foley2015_progenitor_14dt}, 3: \citet{Ochner2014ATel.6648....1O}, 4: \citet{Stahl_2020MNRAS.492.4325S},}} \\
\multicolumn{8}{l}{{\phantom{$^{b}$} 5: \citet{Kawabata2018PASJ...70..111K}, 6: \citet{Singh2018MNRAS.474.2551S}, 7: \citet{Lyman_10.1093/mnras/stx2414}}}
\endlastfoot

+11 & 2014-10-31 & 56961.2 & Copernico/AFOSC & 300 & 450 & \it V & 3 \\
+21 & 2014-11-10 & 56971.84 & Kanata/HOWPol & 900-1200 & 400 & \it V, R, I & 5 \\
+22 & 2014-11-11 & 56972.98 & HCT/HFOSC & 600-1800 & 1330-2190 & \it V & 6 \\
+28 & 2014-11-18 & 56979.0 & HCT/HFOSC & 600-1800 & 1330-2190 & \it V & 6 \\
+29 & 2014-11-18 & 56979.56 & Lick/Kast & 600 & 580-950 & \it V & 2 \\
+32 & 2014-11-21 & 56982.66 & Keck/DEIMOS & 360 & 1400 &  \it V & this paper \\
+34 & 2014-11-23 & 56984.97 & HCT/HFOSC & 600-1800 & 1330-2190 & \it V & 6 \\
+36 & 2014-11-25 & 56986.56 & Lick/Kast & 600 & 580-950 & \it V & 4 \\
+40 & 2014-11-29 & 56990.88 & Kanata/HOWPol & 900-1200 & 400 & \it R, I & 5 \\
+41 & 2014-12-01 & 56992.0 & HCT/HFOSC & 600-1800 & 1330-2190 & \it V & 6 \\
+45 & 2014-12-04 & 56995.93 & HCT/HFOSC & 600-1800 & 1330-2190 & \it V & 6 \\
+48 & 2014-12-08 & 56999.0 & HCT/HFOSC & 600-1800 & 1330-2190 & \it I & 6 \\
+51 & 2014-12-10 & 57001.94 & HCT/HFOSC & 600-1800 & 1330 & \it V & 6 \\
+60 & 2014-12-19 & 57010.91 & HCT/HFOSC & 600-1800 & 1330-2190 & \it V & 6 \\
+61 & 2014-12-20 & 57011.66 & Keck/DEIMOS & 1350 & 1400 &  \it R & this paper \\
+64 & 2014-12-23 & 57014.5 & Lick/Kast & 600 & 580-950 & \it B, V, R, I & 4 \\
+64 & 2014-12-23 & 57014.86 & Kanata/HOWPol & 900-1200 & 400 & \it V, R, I & 5 \\
+66 & 2014-12-26 & 57017.01 & HCT/HFOSC & 600-1800 & 1330 & \it B & 6 \\
+76 & 2015-01-05 & 57027.03 & HCT/HFOSC & 600-1800 & 1330-2190 & \it V & 6 \\
+78 & 2015-01-06 & 57028.84 & HCT/HFOSC & 600-1800 & 1330-2190 & \it V & 6 \\
+82 & 2015-01-10 & 57032.84 & Kanata/HOWPol & 900-1200 & 400 & \it R & 5 \\
+90 & 2015-01-18 & 57040.93 & HCT/HFOSC & 600-1800 & 1330-2190 & \it V & 6 \\
+92 & 2015-01-20 & 57042.51 & Lick/Kast & 1800 & 570-980 & \it V, R, I & 4 \\
+95 & 2015-01-24 & 57046.08 & SALT/RSS & 2200 & 1000 &  \it V & this paper \\
+97 & 2015-01-25 & 57047.81 & HCT/HFOSC & 600-1800 & 1330-2190 & \it V & 6 \\
+100 & 2015-01-28 & 57050.46 & Lick/Kast & 1200 & 580-950 & \it R & 4 \\
+104 & 2015-02-01 & 57054.77 & Kanata/HOWPol & 900-1200 & 400 & \it R & 5 \\
+104 & 2015-02-01 & 57054.86 & HCT/HFOSC & 600-1800 & 1330-2190 & \it V & 6 \\
+108 & 2015-02-05 & 57058.93 & HCT/HFOSC & 600-1800 & 1330-2190 & \it V & 6 \\
+117 & 2015-02-14 & 57067.53 & Lick/Kast & 1800 & 580-950 & \it V & 4 \\
+119 & 2015-02-17 & 57070.03 & SALT/RSS & 1550 & 1000 &  \it R & this paper \\
+124 & 2015-02-21 & 57074.54 & Lick/Kast & 1800 & 580-950 & \it B, V, R, I & 4 \\
+124 & 2015-02-22 & 57075.55 & Subaru/FOCAS & 1200 & 650 & \it R & 5 \\
+126 & 2015-02-23 & 57076.86 & HCT/HFOSC & 600-1800 & 1330-2190 & \it V & 6 \\
+133 & 2015-03-02 & 57083.67 & Kanata/HOWPol & 900-1200 & 400 & \it R & 5 \\
+133 & 2015-03-03 & 57084.01 & SALT/RSS & 1800 & 1000 &  \it V & this paper \\
+138 & 2015-03-08 & 57089.03 & SALT/RSS & 1650 & 1000 &  \it V & this paper \\
+140 & 2015-03-10 & 57091.0 & SALT/RSS & 1600 & 1000 &  \it V & this paper \\
+143 & 2015-03-12 & 57093.79 & HCT/HFOSC & 600-1800 & 1330-2190 & \it V & 6 \\
+148 & 2015-03-18 & 57099.42 & Lick/Kast & 1800 & 580-950 & \it V & 4 \\
+148 & 2015-03-18 & 57099.44 & Keck/DEIMOS & 1800 & 1400 &  \it R & this paper \\
+151 & 2015-03-21 & 57102.55 & Kanata/HOWPol & 900-1200 & 400 & \it R & 5 \\
+156 & 2015-03-25 & 57106.73 & HCT/HFOSC & 600-1800 & 1330-2190 & \it V & 6 \\
+156 & 2015-03-26 & 57107.47 & Lick/Kast & 1800 & 570-980 & \it V & 4 \\
+159 & 2015-03-28 & 57109.9 & SALT/RSS & 1810 & 1000 &  \it V & this paper \\
+160 & 2015-03-29 & 57110.92 & SALT/RSS & 1850 & 1000 &  \it V & this paper \\
+169 & 2015-04-08 & 57120.48 & FTN/FLOYDS & 3600 & 300-600 &  \it B, V, R, I & this paper \\
+171 & 2015-04-10 & 57122.14 & SOAR/GTHS & 1200 & 800 &  \it V & this paper \\
+175 & 2015-04-13 & 57125.93 & SALT/RSS & 1800 & 1000 &  \it B, V & this paper \\
+177 & 2015-04-16 & 57128.41 & Lick/Kast & 1800 & 610-910 & \it R & 1,4 \\
+180 & 2015-04-18 & 57130.9 & SALT/RSS & 1600 & 1000 &  \it V, R, I & this paper \\
+180 & 2015-04-19 & 57131.49 & FTN/FLOYDS & 3600 & 300-600 &  \it V & this paper \\
+184 & 2015-04-22 & 57134.89 & SALT/RSS & 1372 & 1000 &  \it V, R, I & this paper \\
+185 & 2015-04-24 & 57136.6 & FTS/FLOYDS & 3600 & 300-600 &  \it V & this paper \\
+187 & 2015-04-26 & 57138.85 & SALT/RSS & 1525 & 1000 &  \it V & this paper \\
+208 & 2015-05-17 & 57159.83 & SALT/RSS & 1700 & 1000 & \it B, V, R, I & 1 \\
+211 & 2015-05-20 & 57162.39 & FTN/FLOYDS & 3600 & 300-600 &  \it V & this paper \\
+217 & 2015-05-26 & 57168.29 & Lick/Kast & 1800 & 570-980 & \it B, V, R, I & 1,4 \\
+233 & 2015-06-11 & 57184.72 & SALT/RSS & 1700 & 1000 & \it B, V, R, I & 1 \\
+237 & 2015-06-16 & 57189.07 & SOAR/GTHS & 3600 & 800 & \it V & 1 \\
+238 & 2015-06-16 & 57189.33 & Keck/LRIS & 600 & 1000-1700 & \it B, V, R, I & 1,4 \\
+243 & 2015-06-21 & 57194.26 & Subaru/FOCAS & 1800 & 650 & \it V, R, I & 5 \\
+275 & 2015-07-24 & 57227.04 & SOAR/GTHS & 3600 & 800 & \it V & 1 \\
+410 & 2015-12-06 & 57362.59 & Subaru/FOCAS & 1800 & 650 & \it R & 5 \\
+415 & 2015-12-11 & 57367.64 & Keck/LRIS & 2400 & 1000-1700 & \it V & 1 \\
+420 & 2015-12-16 & 57372.59 & Keck/DEIMOS & 1200 & 1400 &  \it R, I & this paper \\
+454 & 2016-01-20 & 57407.0 & VLT/MUSE & 2080 & 2000-4000 & \it R, I & 7 \\
+456 & 2016-01-22 & 57409.0 & Gemini/GMOS-N & 7200 & 640 &  \it R & this paper \\
+562 & 2016-05-07 & 57515.0 & Keck/LRIS & 1200 & 1100-1700 & \it R & 4 \\

\end{longtable*}

After reduction or collection, all spectra went through the following series of steps: (1) manual cleaning, (2) scaling/mangling, (3) Galactic extinction correction, and (4) host-galaxy redshift correction. First, we manually identified and removed any remaining cosmic rays or telluric features, and masked narrow interstellar emission lines from the host galaxy. Next, for absolute flux calibration, we ``mangled'' the spectra to match broadband optical \textit{BVRI} photometry of SN\,2014dt from \citet{Singh2018MNRAS.474.2551S}, \citet{Kawabata2018PASJ...70..111K}, and Johansson et al. (in prep.). The photometry is fit via a Gaussian Process (GP) model that produces smooth, continuous \textit{BVRI} light curves with uncertainties that account for data uncertainties and scatter. In the mangling procedure, we multiply the spectra by a smooth function in wavelength so that the passband-integrated flux matches the value of the corresponding GP light curve at that time. Some of the spectra had enough wavelength coverage to match multiple optical filters; for these spectra we verified that the mangling did not significantly change the relative line strengths -- generally, the relative flux adjustments were no more than $\sim 20$\%. For a few spectra where the mangling procedure unacceptably altered the spectrum (most likely owing to uncertainties in the photometry or background contamination in the spectra), we only apply a uniform rescaling to one filter. Spectra that only had enough wavelength coverage to match one optical filter were rescaled to that single photometric band. The filters used to scale/mangle each spectrum are listed in \autoref{tab:obs}. 

For all spectra, we next correct for Milky Way extinction $E(B-V) = 0.02$ mag with $R_V = 3.1$ \citep{Schlafly2011ApJ...737..103S}. We did not further correct for host-galaxy extinction, thought to be negligible for SN\,2014dt \citep{Singh2018MNRAS.474.2551S,Foley2015_progenitor_14dt}. Lastly, we transform the spectra to the host-galaxy rest frame, correcting for the redshift of M61, $z = 0.0052$ \citep[via the NASA/IPAC Extragalactic Database, NED]{Allison:2014}.

In total, we present 69 new and previously published SN\,2014dt spectra, making SN\,2014dt one of the best spectroscopically observed SN. Details of the observations, including phase, date, telescope+instrument, exposure time, spectral resolution, and which filter(s) were used for scaling/mangling can be found in \autoref{tab:obs}. The full spectral evolution of SN\,2014dt -- after cleaning, scaling/mangling, Galactic extinction correction, and host-galaxy redshift correction -- is shown in \autoref{fig: all_spec_14dt} and is available as data behind the figure. New spectra, denoted with $\star$ in \autoref{fig: all_spec_14dt}, will be available on \href{https://wiserep.weizmann.ac.il}{WISeREP}. 

Our spectral modelling in \autoref{sec:synthetic_spec} requires the additional step of converting our spectra from flux to luminosity units by using the distance to M61. Distances to M61 from the NASA/IPAC Extragalactic Database (NED) range from 7.6 Mpc \citep{1984A&AS...56..381Bottinelli} to 35.5 Mpc \citep{1994ApJ...433...19Sparks} with a mean distance of 14.6 Mpc. Recent papers on SN\,2014dt choose varying distances around this value. In particular, \citet{Foley2015_progenitor_14dt, Foley2016} adopt 12.3 Mpc, \citet{2016ApJ.Fox...816L..13F} use 19.3 Mpc, \citet{Singh2018MNRAS.474.2551S} adopt 21.4 Mpc, and \citet{Kawabata2018PASJ...70..111K} use 14.5 Mpc. This range of distances would produce a difference of $\sim 1$ mag for the absolute maximum magnitude in \textit{B}, $M^{B}_{\rm max}$, and correspondingly change the luminosity we obtain from our spectra. In this paper we adopt a distance of 14.6 Mpc, but we also explore the effects of a range of distances from 12.3 to 21.4 Mpc. As we describe in \autoref{sec: spec_results_analysis}, distances at the lower end of this range have a minimal impact on our spectral modelling results, while higher assumed distances produce larger changes in our modelling that are less compatible with the observations. Our analysis favors a distance to SN\,2014dt less than 18 Mpc.

\begin{figure*}[htb]
\centering
\includegraphics[width=\textwidth]{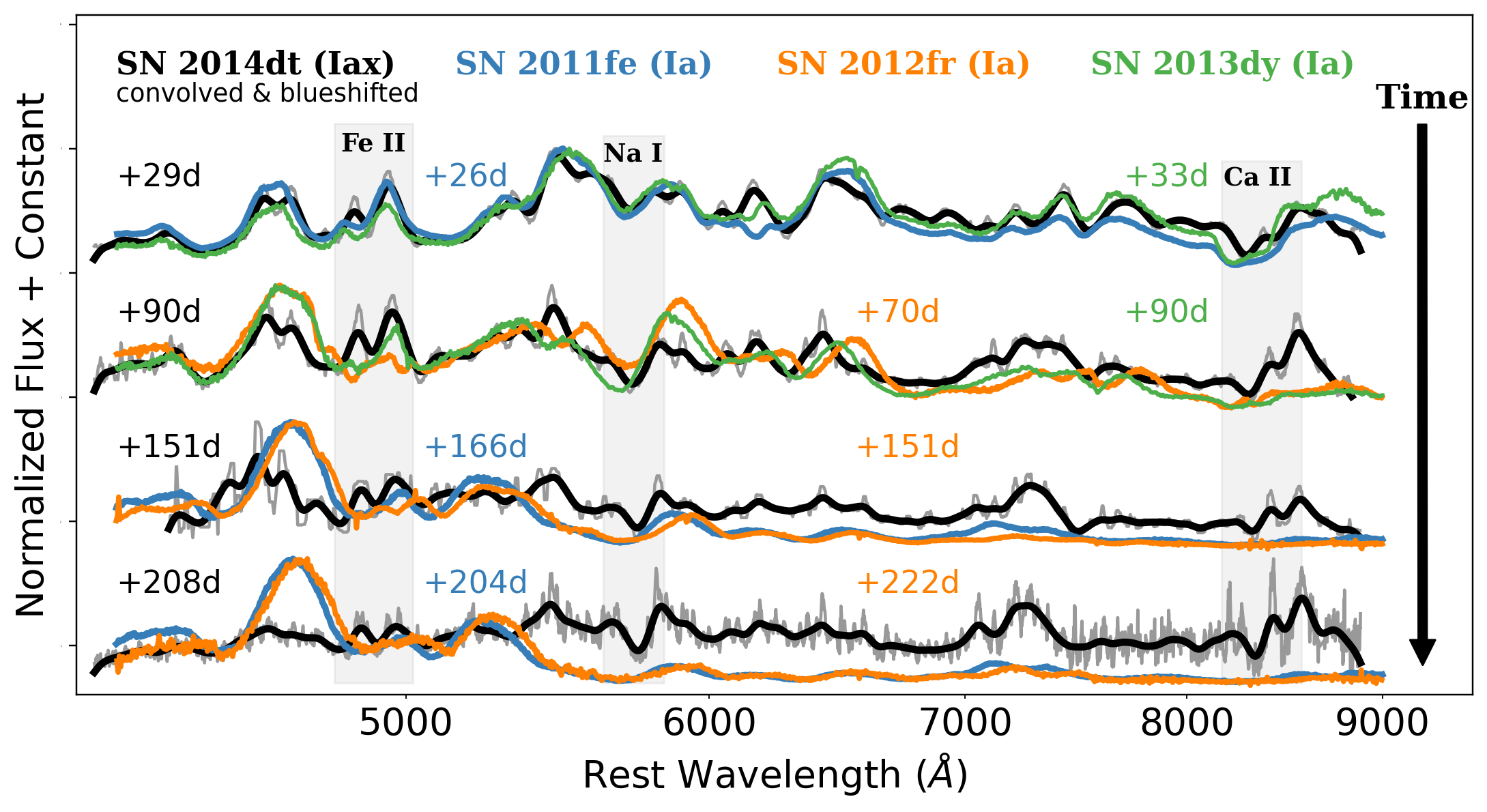}
\caption{The spectrum of Type Iax SN\,2014dt (black/gray) diverges from the spectrum of normal Type Ia SN (blue, orange, green) over time. The black SN\,2014dt spectrum has been convolved with a velocity of 1200 km s$^{-1}$ and blueshifted by 4000 km s$^{-1}$ in order to compare its spectral features with the higher velocities of normal SN Ia. The unconvolved, but still blueshifted, SN\,2014dt spectrum is shown in gray. All spectra of normal SN Ia were collected from The Open Supernova Catalog \citep[][]{2017ApJ_opensupernovacatalog...835...64G} and came from the following sources: SN\,2011fe \citep{Mazzali2014MNRAS, 2012PASP.Ofer..124..668Y,Stahl_2020MNRAS.492.4325S,Silverman_2012MNRAS.425.1789S, 2015MNRAS.Mazzali.450.2631M}, SN\,2012fr \citep{2012PASP.Ofer..124..668Y, 2012fr_2014_Zhang, 2015MNRAS.Childress.454.3816C}, and SN\,2013dy \citep{Stahl_2020MNRAS.492.4325S, Silverman_2012MNRAS.425.1789S}.}
\label{fig: Iax_vs_normalIa}
\end{figure*}

\section{Spectral Divergence Phase}\label{sec:diverg}

As discussed in \autoref{sec:intro}, SN Iax are spectroscopically similar to normal SN Ia at early times, but by late times their spectra are much different \citep{Jha2006, Sahu2008, Foley2010ApJ, Foley2016}. The extensive spectral coverage of SN\,2014dt between +80 and +200 days allows us to pinpoint the phase when it transitions away from normal SN Ia spectra, which we call the spectral divergence phase of SN Iax.

In \autoref{fig: Iax_vs_normalIa}, the spectral evolution of SN Iax 2014dt (black/gray) is compared with the spectral evolution of other normal SN Ia (colors) from $\sim$ +30 days to $\sim$ +200 days. In order to account for the lower velocity of SN\,2014dt, we convolve its spectra with a Gaussian with a full width at half-maximum intensity (FWHM) of 1200 km s$^{-1}$ and blueshift by 4000 km s$^{-1}$ before comparing it to the higher velocity, normal SN Ia. The early epoch (+29 days) spectrum of SN\,2014dt shows remarkable similarity with normal SN Ia spectra for all features at optical wavelengths. By $\sim$ +90 days, the prominent Fe~II double-peak feature at $\sim 5000$ \AA\ starts to disappear in normal SN Ia but remains present in SN\,2014dt until very late epochs (disappearing sometime between +275 and +410 days; see \autoref{fig: all_spec_14dt}). The Ca~II near-infrared triplet $\lambda\lambda 8498$, 8542, 8662 (hereafter Ca~II IR triplet) in normal SN Ia is also nearly gone by $\sim$ +90 days and completely gone by $\sim$ +150 days, but continues to be a strong feature in SN\,2014dt. The broad Na~I~D $\lambda$5890 \AA\ absorption is strong in SN\,2014dt at all epochs, but starts to disappear in normal SN Ia by $\sim$ +150 days. In general, by +150 days the spectra of normal SN Ia lose most of their photospheric (permitted) features and are dominated by broad, nebular (forbidden) emissions of [Fe~II] and [Fe~III], while the spectra of SN\,2014dt continue to be dominated by photospheric features and remain largely unchanged after +90 days. For these reasons, we conclude that the spectral divergence phase of SN\,2014dt lasts $\sim 60$ days between +90 days and +150 days past $B_{\rm max}$.

\section{Spectral Models and Analysis}\label{sec:synthetic_spec}

Attempts at modelling the early-time ($<$ +41 days) spectra of SN\,2014dt have been carried out by \citet{Singh2018MNRAS.474.2551S} and \citet{Kawabata2018PASJ...70..111K}. These early-time models were used primarily for line identification and show prominent, often blended, lines of IMEs and IGEs; most notably Fe~II, Co~II, and Ca~II. We present spectral modelling that spans from +11 days, 10 days earlier than the earliest spectra previously modelled, to +562 days, 521 days later than the latest spectra previously modelled. Our early-time models generally agree with the line identifications found in these previous studies and offer additional insights into the evolving contributions from different elements and ions. 

We use \tardis\ \citep{Kerzendorf2014MNRAS.440..387K,kerzendorf_wolfgang_2023_7525913}, a 1D open-source Monte Carlo (MC) radiative transfer code to model the spectra of SN\,2014dt. It generates model spectra based on simple assumptions: spherical symmetry, homologous expansion ($r = vt$), and a sharp photosphere that emits a blackbody spectrum (i.e., it cannot generate forbidden-line emission). Indivisible MC packets representing bundles of photons with the same frequency are followed as they go from the photosphere through the ejecta, until they escape to create the model spectrum. \tardis\ has been used to model many different kinds of supernovae, including other SN Iax: SN\,2002cx \citep{Barna2018_mnras}, SN\,2005hk \citep{Barna2018_mnras, Magee2017AA...601A..62M, Magee_2019}, SN\,2007J \citep{Magee_2019}, SN\,2010ae \citep{Magee_2019}, SN\,2011ay \citep{2017MNRAS.Barna.471.4865B, Barna2018_mnras}, SN\,2012Z \citep{Barna2018_mnras, Magee_2019, magee2022MNRAS.509.3580M}, PS1-12bwh \citep{Magee2017AA...601A..62M}, SN\,2015H \citep{Magee2016A&A, Barna2018_mnras, Magee_2019}, SN\,2019muj \citep{2020Barna}, SN\,2019gsc \citep{Srivastav2020ApJ...892L..24S}, and SN\,2020sck  \citep{2022ApJ.Dutta...925..217D}.

\begin{figure*}[htp!]
\includegraphics[width=\textwidth]{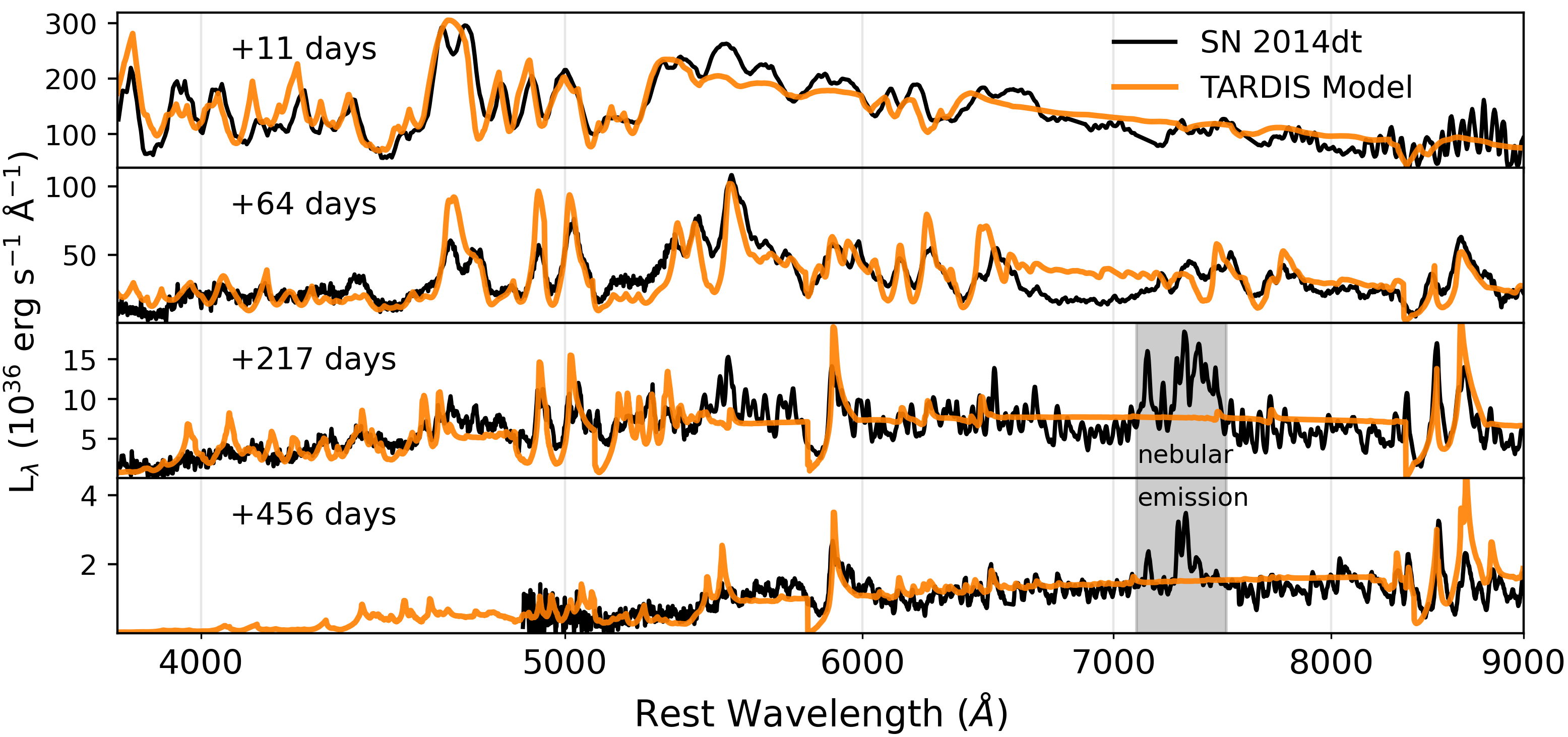}
\caption{Model spectra (orange) from \tardis\ for SN\,2014dt (black) at +24, +59, +212, and +415 days past $B_{\rm max}$. Most features and line widths agree with observations. Nebular emission at later epochs cannot be modeled by \tardis\ (gray region).}
\label{fig: best_TARDIS_fits}
\end{figure*}

\tardis\ is typically only able to model relatively early-time spectra (less than a few weeks after peak magnitude) because the assumption of a sharp photosphere starts to break down at this time for most supernovae. However, since SN Iax spectra never become dominated by forbidden-line emission, we are able to model the optical spectra of SN\,2014dt out to extremely late times. We construct \tardis\ models for all SN\,2014dt spectra from +11 days to +562 days past $B_{\rm max}$ with generally good to excellent fits for all epochs. In \autoref{fig: best_TARDIS_fits}, a few early, mid, and late epochs are chosen to demonstrate how the model spectra (generated using the formal integral method from \tardis\ to reduce the MC noise) and observed spectra of SN\,2014dt compare. The model spectra correctly recreate the continuum, spectral lines, and relative strengths of most of the photospheric features in the observed spectra. When deciding on the best-fit model, we tried to match the full wavelength range of each spectrum while emphasizing a good match to the blue end of the spectrum in the range $\sim 3500$--5500 \AA, a region with many permitted lines of IGEs. In the latest two epochs in \autoref{fig: best_TARDIS_fits}, +217 and +456 days past $B_{\rm max}$, the gray region includes nebular emission at $\sim 7200$ \AA; these nebular regions were excluded during our fitting since \tardis\ is not able to model these forbidden Fe, Ca, and Ni lines \citep{Foley2016}. 

The input parameters for each model spectrum are discussed subsequently in \autoref{sec: spec_inputs}. A detailed analysis of the resulting synthetic spectra follows in \autoref{sec: spec_results_analysis}, including a closer look at the individual contributions from different ions and elements.

\begin{figure*}[htbp!]
\centering
\includegraphics[width=\linewidth]{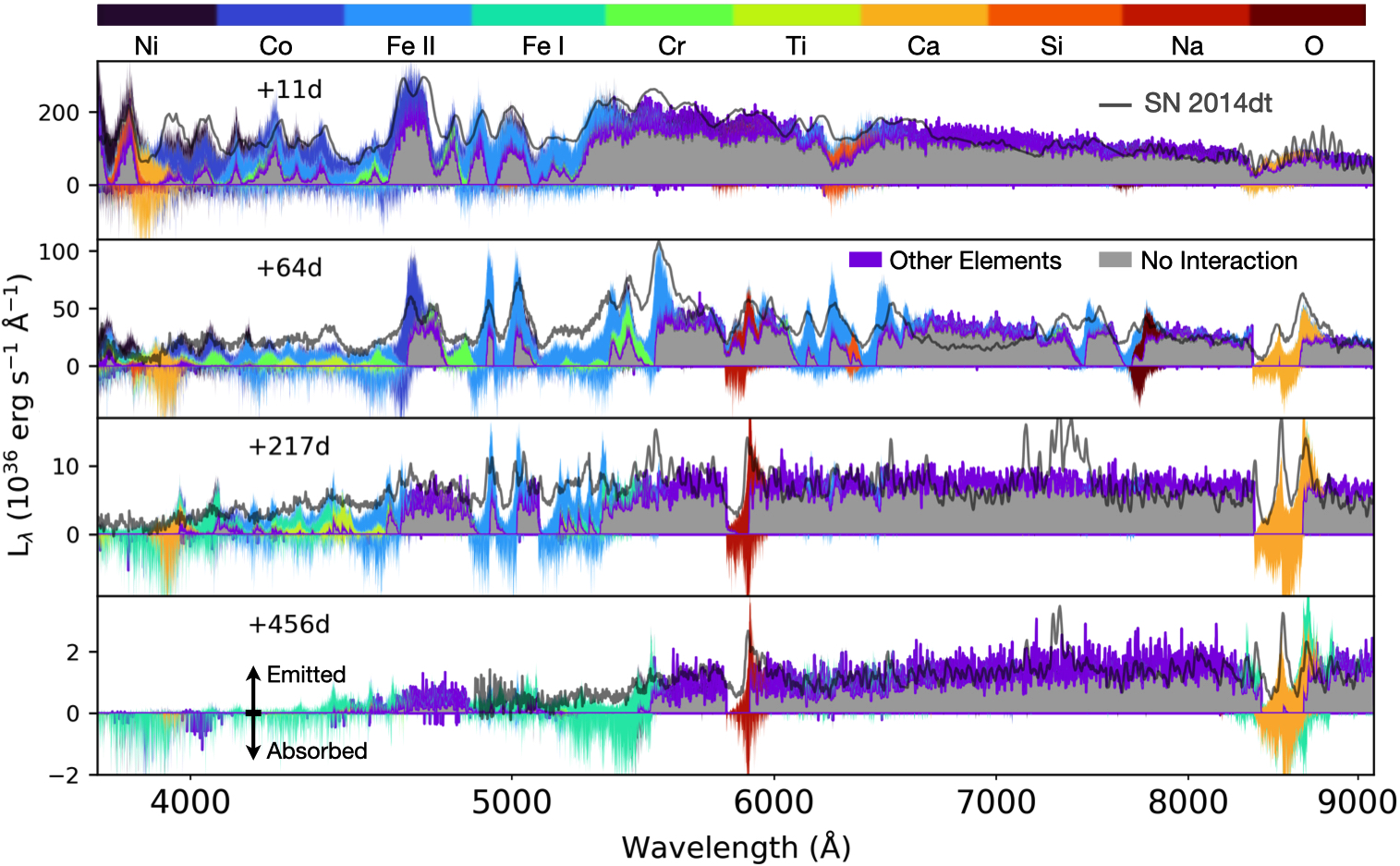}
\caption{Element decomposition of model spectra at +11, +64, +217, and +456 days past $B_{\rm max}$. SN\,2014dt (dark-gray line) is shown for comparison. Elements and ions with the greatest contributions are shown on the color map. Contributions from all other elements are in purple and packets that did not interact with the ejecta (blackbody emission from the assumed photosphere) are in light gray. Emission is shown as positive (above zero luminosity) and absorption is negative.}
\label{fig: SDEC_evolution_full}
\end{figure*}

\subsection{Synthetic Spectra: Input Parameters}\label{sec: spec_inputs}

\tardis\ needs a number of input parameters in order to generate each synthetic spectrum. Some inputs are used to set up the computational framework (e.g., number of iterations, number of MC packets) for each simulation. Of particular importance, the modes used to calculate the state and behavior of plasma for all of our simulations are 
\begin{itemize}[leftmargin=10pt,noitemsep]
    \item[--] \texttt{ionization: nebular}
    \item[--] \texttt{excitation: dilute-lte}
    \item[--] \texttt{radiative$\_$rates$\_$type: dilute-blackbody}
    \item[--] \texttt{line$\_$interaction$\_$type: macroatom \ .}
\end{itemize} 
\noindent
Other input parameters pertaining to our specific SN include the luminosity requested, time since explosion, chemical abundance profile, density profile, and inner (\vphot), and outer (\vouter) boundaries of the line-forming region; details can be found in \autoref{tab: model_parameters}. All \tardis\ input files and output synthetic spectra are available on \href{https://github.com/ycamacho/SN-2014dt-TARDIS}{GitHub}\footnote{\url{https://github.com/ycamacho/SN-2014dt-TARDIS}} and \dataset[doi:10.5281/zenodo.7838690]{https://doi.org/10.5281/zenodo.7838690}.

The luminosity, $L_{\rm{requested}}$, comes directly from integrating each observed spectrum. The $L$ integration range, $\lambda_{\rm{min}}$ -- $\lambda_{\rm{max}}$, is based on the observation limits, sometimes restricted to exclude noisy regions at the ends of a spectrum.

We set the time since explosion to be equal to rest-frame days past $B_{\rm max}$, plus 11 days to account for the rise time from explosion to peak. Unfortunately, SN\,2014dt was found after peak brightness and does not have data for a precise rise-time measurement. \citet{Kawabata2018PASJ...70..111K} estimate a wide range of possible rise times, 11.8 to 21.0 days for the bolometric light curve of SN\,2014dt (which may differ slightly from the rise time to $B_{\rm max}$ that we desire). Our choice of an 11 day rise time is at the lower end, but roughly consistent with their results. It is also within the $\sim 10$ to 20 days rise time typical for SN Iax \citep{Jha2017}, and it is in the range of rise times found in the relevant \citet{Fink2014MNRAS} deflagration models that we use. For completeness, we explore a range of rise times up to 20 days, and find that our spectral-fitting results are not very sensitive to choice of rise time, particularly for epochs after about +40 days. Our choice of rise time gives the best fits to the earlier epochs, but different choices are not definitively worse.

Remarkably, we are able to use a self-consistent description of the ejecta (fixed chemical abundance and density profiles as a function of velocity) to generate model spectra at the wide range of epochs covered by the data. As mentioned in \autoref{sec:intro}, deflagration models are generally favored for SN Iax. We model the spectra of SN\,2014dt using numerous deflagration models from \citet{Fink2014MNRAS}. These models are 3D deflagration explosions of a Chandrasekhar-mass carbon-oxygen WD. In order to be used by the 1D \tardis\ code, we obtain the spherically averaged density and isotopic abundance profiles from the Heidelberg Supernova Model Archive\footnote{\url{https://hesma.h-its.org/doku.php}} \citep[HESMA;][]{Kromer2017MmSAI..88..312K}. 

We make three small adjustments to the \citet{Fink2014MNRAS} model isotope abundances. First, we remove all elements past Zn, the atomic-data limit of \tardis. The total abundance (mass fraction) of removed elements is small, $\sim 10^{-6}$, and is not expected to have a noticeable effect on the model spectrum. Second, we use a uniform, averaged isotope abundance value for all shells. We find that this has little to no impact on the resulting spectrum since the model ejecta are already well-mixed (i.e., abundances are nearly constant as a function of velocity; see \autoref{sec: comp_deflagration_models} for details of modelling with velocity-dependent abundances). Third, we increase the abundance of $^{23}$Na from $\sim 0.01\%$ to $\sim 1.7\%$ (discussed further in \autoref{sec: resonant_lines}). 

The only modification made to the \citet{Fink2014MNRAS} model density profiles was to extend them below the original tabulation limit of $\sim 200$ km s$^{-1}$ via a cubic spline extrapolation down to zero velocity. This extrapolated velocity range is only relevant for modelling the latest epochs. 

The line-forming region is specified by the inner velocity boundary, \vphot, and the outer velocity boundary, \vouter. In general, the synthetic spectrum is much more sensitive to the choice of \vphot\ than to the choice of \vouter. Our choice of \vouter\ is based on fitting the Ca~II IR-triplet, the main feature affected by \vouter, and is discussed further in \autoref{sec: resonant_lines}. This leaves \vphot\ as our primary fitting parameter for each spectral epoch. 

\subsection{Synthetic Spectra: Results \& Analysis}\label{sec: spec_results_analysis}

We find that using the n1def deflagration model density and abundance from \citet{Fink2014MNRAS} produces the best-fit synthetic spectra for SN\,2014dt. The results for a wide range of epochs are presented in \autoref{fig: best_TARDIS_fits}. The level of agreement between the observed spectra of SN\,2014dt and the synthetic spectra generated by a nearly unaltered deflagration model (as described in \autoref{sec: spec_inputs}) is quite remarkable. The continuum, spectral lines, and relative strengths of most of the photospheric features in the observed spectra are well reproduced by our model for all epochs, covering over 500 days of evolution.

\autoref{fig: SDEC_evolution_full} shows the observed spectra in gray compared to the synthetic spectra, which are broken down and color-coded by contributions from different elements and ions. Contributions come from each photon packet's last interaction, with emission shown as positive luminosity and absorption as negative luminosity. Some wavelength regions in each spectrum are dominated by the underlying blackbody photosphere (labeled ``No Interaction''), while other regions have significant contributions from one or many different elements and ions. In particular, the blue end ($< 5500$ \AA) of spectra up to +217 days exhibits complex line contributions. Using both \autoref{fig: best_TARDIS_fits} and \autoref{fig: SDEC_evolution_full}, we now discuss the details of how well the synthetic spectra reproduce the observed spectra of SN\,2014dt.

In the earliest epoch, +11 days, the region between $\sim 3900$ and $\sim 5500$ \AA\ is particularly well matched with almost all the spectral features present. This is a blended region of IGEs, dominated by permitted lines of Co and Fe with some contributions from Ni and Cr. The redder end ($>5500$ \AA) of the +11 day spectrum is also mostly reproduced by our model, with contributions from Ca~II, O~I, and Si~II.

The +64 day synthetic spectrum reproduces most of the photospheric features, and matches the Ca~II IR-triplet, as well as the Fe~II features at $\sim 5000$ \AA\ and $\sim 6200$ \AA\ particularly well. By +217 days, while the relative strengths of the photospheric lines have decreased as the nebular emission at $\sim 7200$ \AA\ increases, many photospheric lines continue to be reproduced well by our model, especially Fe~II $\sim 5000$ \AA\ and the Ca~II IR-triplet. More than 200 days later, the +456 day spectrum continues to match the now Fe~I features at $\sim 5500$ \AA\ and the Ca~II IR-triplet. 

Although the model allows us to explore the contribution of each ionization state for each element, for most elements, one ionization state dominates all epochs, e.g., O I, Na I, Si II, Ca II. For IGEs, multiple ionization states are relevant. We group together these contributions for Cr, Co, and Ni for clarity in \autoref{fig: SDEC_evolution_full}, but show separately the contributions of Fe II and Fe I to illustrate how the ionization evolves with time. In the earliest epochs, +11 and +64 days, Fe II dominates the hot ejecta. As the ejecta cool and expand with time, we see the contribution from Fe I increase and by +217 days it makes up most of the Fe at $< 4500$ \AA\ while Fe II still dominates at $\sim 5000$ \AA. By +456 days, all the Fe lines in the optical regime come from Fe I. 

For the +11 day epoch, while the relative line strengths and location of prominent features are well matched, there are some features that appear slightly too blueshifted when compared with the observed spectrum. In particular, the model line profiles around Si II $\lambda$6355 \AA\ shows a blueshift $\sim 1000$ km s$^{-1}$ more than expected from observations. This might be due to too much Si in the faster, outer layers of the ejecta of our model or some other asymmetry in the ejecta structure or explosion propagation. We have assumed uniform abundance profiles for all elements, supported by \citet{2017MNRAS.Barna.471.4865B} and \citet{magee2022MNRAS.509.3580M}, but their work focuses on the outer layers around maximum light, and it is nonetheless possible the ejecta of SN Iax require some degree of stratification. 

One notable missing feature in the +11 to +64 day spectra is the absorption at $\sim 4670$ \AA. This exact absorption feature was also not recovered in the very similar +10.4 day SN Iax 2015H spectrum that was modelled by \citet{Barna2018_mnras}. Both \citet{Kawabata2018PASJ...70..111K} and \citet{Singh2018MNRAS.474.2551S} are able to reproduce the appearance, but not the correct strength, of this absorption feature in their model SN\,2014dt spectra, but the former credits the feature to Fe II and the latter thinks it might be [Fe III]. Although we are unable to produce this feature with the n1def model, we were able to recreate it with the n20def model for only the earliest epoch, +11 days. This model showed that the absorption at $\sim 4670$ \AA\ is very blended with contributions from many IGEs and IMEs -- the most notable difference being more C II interaction and less contribution from the photospheric blackbody. However, besides the $\sim 4670$ \AA\ absorption, the n20def model does not match the earliest spectra of SN\,2014dt nearly as well as the n1def model, and is unable to reproduce this absorption feature for any later epoch. The identity of the prominent absorption at $\sim 4670$ \AA\ therefore remains unclear and is likely a complex blended line.

As described at the end of \autoref{sec:obs}, we adopted a distance of 14.6 Mpc for SN\,2014dt to calculate the luminosity input parameter, $L_{\rm{requested}}$. Because of the large uncertainty in the distance to M61 we also explored a wide range of distances in the range 12.3--21.4 Mpc. We find that the late-time spectra, after about +64 days, are insignificantly affected by changes in the assumed distance, so we focus our varying-distance analysis on earlier epochs. We find that using distances below 14.6 Mpc has a small adverse effect on the overall fit of the spectrum produced that can be easily remedied by using a slightly lower model \vphot\ value. However, using distances at the higher end of the plausible range produced poor fits to the observed spectra, even when we tested a wide range of \vphot\ values. In those cases, the best \vphot\ value to match the spectral features produced a continuum that was too hot and blue, and the best \vphot\ value that matched the continuum ($\sim 2500$ km s$^{-1}$ higher than our original \vphot) produced a featureless spectrum beyond $\sim 5200$ \AA\ and an overall poor fit to the blue end under $\sim 5200$ \AA. Based on the results of this analysis, along with the remarkable agreement of the n1def model to the spectra of SN\,2014dt when using shorter distances, we suggest that the correct distance to M61 is likely less than about 18 Mpc.

\citet{Maeda_2022} recently analyze the Type Iax SN~2019muj with spectral modelling similar to that presented here. They find evidence for a dense inner core of a few hundredths of a solar mass, in contrast with the density profiles of the deflagration models we use. We find that we are able to satisfactorily model SN\,2014dt without a dense inner core. We tested the \citet{Maeda_2022} inner core density profile for SN\,2014dt and find that it produces reasonable fits to the early-time spectra, $\lesssim$ +100 days, but the n1def density profile without a central density enhancement still produces better fits. Fitting the late-time spectra with the dense inner core model requires a higher \vphot\ than in our n1def model and does not appear to improve the overall fit to the late-time spectra. Thus, while we cannot rule out an inner core in SN~2014dt, we do not require one. Despite these different inner density profiles, we find remarkable agreement in the photospheric velocities of SN~2014dt and those measured for SN~2019muj by \citet{Maeda_2022}, even at the latest epochs at $\sim$ +500 days. 

\subsection{Comparing Deflagration Models}\label{sec: comp_deflagration_models}

\begin{figure*}[htbp!]
\includegraphics[width=\linewidth]{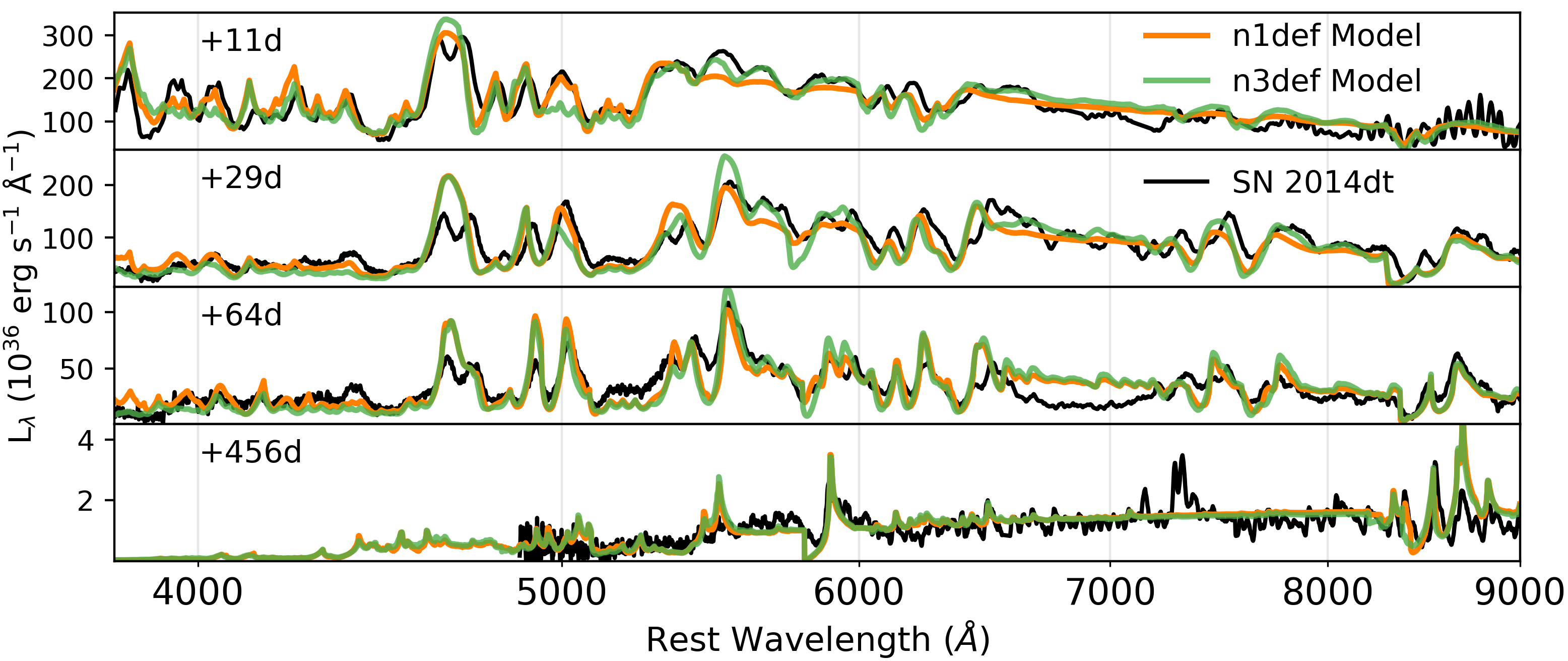}
\vspace{1cm}
\includegraphics[width=\linewidth]{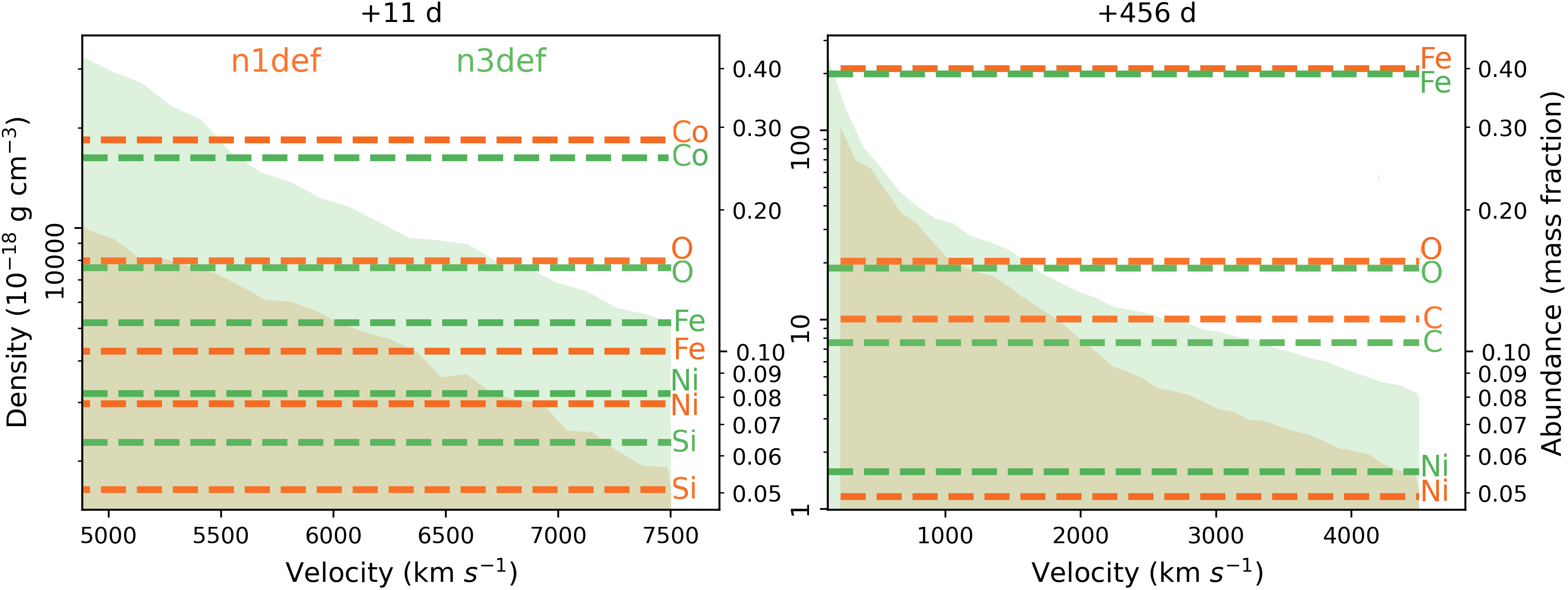}
\caption{\textit{Top panel:} Early-time and late-time \tardis\ spectra using the n1def (orange) and n3def (green) deflagration models from \citet{Fink2014MNRAS}. Both deflagration models fit the observed spectrum of SN\,2014dt well; early epochs show the biggest differences, favoring the n1def model, and both models become extremely similar after +64 days. \textit{Bottom panels:} Density and select element abundances of the n1def and n3def models at early and late times.
\label{fig:deflagration_model_comp}}
\end{figure*}

\begin{table*}[hbtp!]
\caption{Comparison of SN\,2014dt properties inferred in published work with our best-fit spectral models. The different assumed distances affect the derived absolute magnitude and mass estimates. Details are discussed in \autoref{sec: comp_deflagration_models}.}
\label{tab: def_models_vs_14dt_obs}
\centering
 \begin{tabular}{c|ccccccc} 
 \hline
  & Distance & $M^{B}_{\rm max}$ & $B_{\rm max}$ & $t^{B}_{\rm max}$ & $\Delta m^{B}_{15}$ & $M_{\rm ejecta}$ & $M_{\rm Ni56}$\\  
  & (Mpc) & (mag)  & (mag) & (days)   & (mag)    & (M$_{\odot}$)  & (M$_{\odot}$) \\ [0.5ex] 
 \hline\hline
\citet{Singh2018MNRAS.474.2551S} & 21.44 $\pm$ 0.03 & $-$18.13 $\pm$ 0.04  & 13.59 $\pm$ 0.04 & -- & 1.35 $\pm$ 0.06 & 0.95 & $>$0.14 \\
\citet{Kawabata2018PASJ...70..111K} & 13.24 -- 15.92 & $-$ (16.82 -- 17.34) & 13.76 -- 13.88 & 11.8 -- 21.0$^{\ast}$& 1.43 -- 1.57 & 0.08 -- 0.35 & 0.035 -- 0.105 \\
 \citet{Fink2014MNRAS} n1def model & 14.6$^{\ast\ast}$ & $-$16.55  & 14.3$^{\ast\ast}$ & 7.6 & 2.15 & 0.0843 & 0.0345 \\ 
 \citet{Fink2014MNRAS} n3def model & 14.6$^{\ast\ast}$ & $-$17.16  & 13.7$^{\ast\ast}$ & 9.6 & 1.91 & 0.195 & 0.0730 \\ [1ex] 
 \hline
\multicolumn{5}{l}{{$^{\ast}$ Estimated rise time of bolometric light curve from other SN Iax.}}\\
\multicolumn{5}{l}{{$^{\ast\ast}$ Using our adopted distance, 14.6 Mpc, and Milky Way extinction, $A_V = 0.062$ mag.}}
 \end{tabular}
\end{table*}

Although the n1def model produced synthetic spectra that most closely resembled SN\,2014dt, other deflagration models from \citet{Fink2014MNRAS} with few ignition points (n3def in particular) also do a reasonable job. The density profile and abundances of n3def were generated in the same way as n1def, as described in \autoref{sec: spec_inputs}. We keep all other \tardis\ input parameters the same for each epoch, with only a very small change in the \vphot\ parameter for the latest epochs, $>$ +410 days.

In \autoref{fig:deflagration_model_comp}, we compare the n1def and n3def deflagration models at various epochs. The top panel shows the resulting synthetic spectra while the bottom panel illustrates the density and abundance structure of each model. Though both models produce extremely similar spectra, as expected, the earliest epoch, +11 days, exhibits the greatest difference in spectral features. There are three main regions in the optical spectra that differ: 3900--4100 \AA, 5000 \AA, and 5300--6000 \AA. The best-fit model, n1def, was chosen based on fitting SN\,2014dt better in the 3900--4100 \AA\ and 5000 \AA\ regions. In particular, the peak at 5000 \AA\ is fit extremely well by n1def while n3def is completely unable to produce this prominent feature. By +29 days, the difference between n1def and n3def at 5000 \AA\ has declined but the difference at 3900--4100 \AA\ and 5300--6000 \AA\ remains. By +64 days, both deflagration models produce very similar spectra with only small differences, mostly $< 4100$ \AA. This similarity continues for all subsequent epochs, as shown by the +456 days spectrum.

The bottom panel in \autoref{fig:deflagration_model_comp} compares the density profiles and element abundances of n1def and n3def at early and late times. The velocity range for each epoch is set from \vphot\ to \vouter; note that at late times we probe much deeper (lower-velocity) layers in the ejecta. Both n1def and n3def have similarly declining slopes in their density profiles but n1def always has a lower density at a given velocity. The abundance profiles of both models are very similar and we select a few elements with high abundances to show in \autoref{fig:deflagration_model_comp}. N1def has slightly more C and O, while n3def has more Si. Since the C, O, and Si abundances are stable over time, they are not all shown for both epochs. The abundances of elements with radioactive isotopes do change with time -- this radioactive decay is done internally by \tardis. At early times, +11 days, n1def has more Co, while n3def has more Fe. By +456 days, almost all of the radioactive Ni and Co has decayed to Fe. The abundance of Co becomes so small that it falls below the lower limit of our plot. The remaining Ni shown at +456 days is stable $^{58}$Ni. The abundance of Fe at this late time is high for both models, with n1def having slightly more Fe than n3def.

\begin{figure}[htbp!]
\includegraphics[width=1\linewidth]
{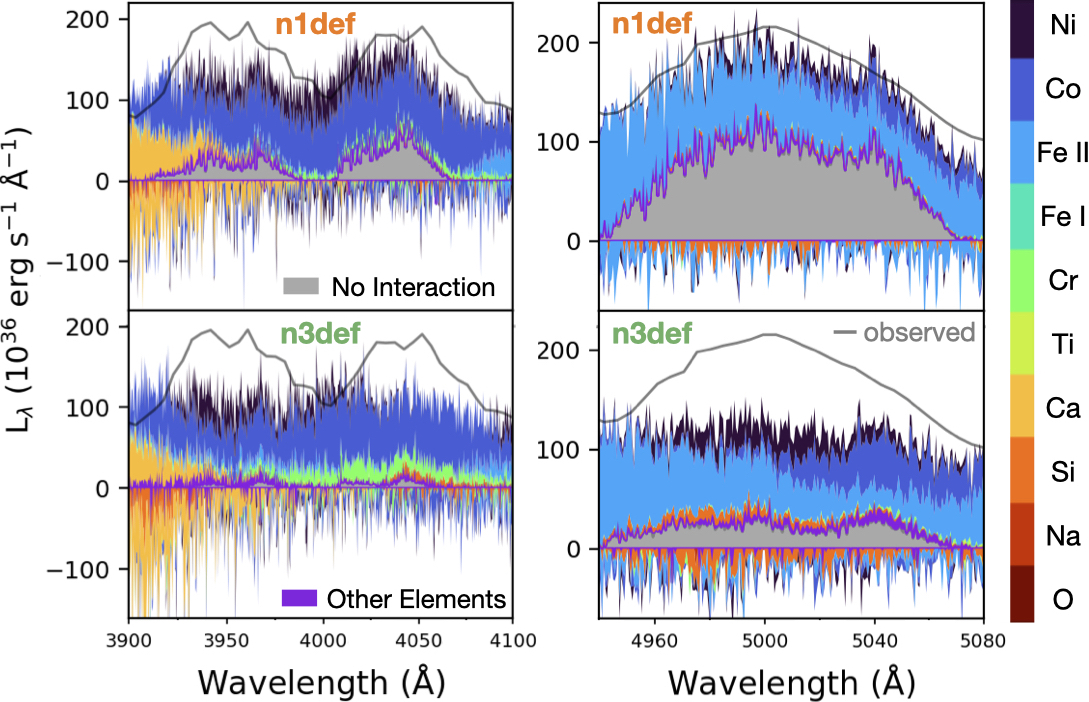}
\caption{Element decomposition of n1def and n3def model spectra at wavelengths where they differ most, compared to the observed SN\,2014dt spectrum at +11 days.
\label{fig: n1def_n3def_sdec4}}
\end{figure}

A closer look at the 3900--4100 \AA\ and 5000 \AA\ regions in \autoref{fig: n1def_n3def_sdec4} shows that more emission from the underlying blackbody (labelled ``No Interaction'') in n1def is one of the reasons why this model matches SN\,2014dt better at early epochs; we note that the blackbody radiation we see in our models is limited by \tardis\ originating photons solely from the photosphere. The lower density of n1def leads, in part, to a lower optical depth at these wavelengths and allows more of the blackbody light through. N1def and n3def show similar, but not identical, contributions from elements and isotopes at these wavelengths; the main differences include n3def having more Cr emission at 3900--4100 \AA, and more Co emission and Si absorption around 5000 \AA.

The differences seen in the early-time spectra of n1def and n3def are mostly due to their different density profiles rather than abundances, such that a model with n1def (n3def) density and n3def (n1def) abundance produces a spectrum that looks like a pure n1def (n3def) model. This sensitivity to the density profile, coupled with both models producing very similar spectra after +64 days, leads us to conclude that the density profile between $\sim 1000$ and 7500 km s$^{-1}$ is a key fitting parameter when reproducing the spectra of SN\,2014dt.

In \autoref{tab: def_models_vs_14dt_obs} we compare the observed properties of SN\,2014dt to those generated by the n1def and n3def models. Owing to a large difference in distances used by \citet{Kawabata2018PASJ...70..111K} and \citet{Singh2018MNRAS.474.2551S}, it is difficult to compare their absolute brightness, $M^{B}_{\rm max}$, so we focus on $B_{\rm max}$ instead. \citet{Kawabata2018PASJ...70..111K} estimate a $B_{\rm max}$ of 13.76--13.88 mag while  \citet{Singh2018MNRAS.474.2551S} find a similar, but not overlapping, 13.59 $\pm$ 0.04 mag. The observed $B_{\rm max}$ is a close match to the n3def model for our adopted extinction and distance of 14.6 Mpc. Even though the n1def model better matches our spectra, it predicts a $B_{\rm max}$ that is slightly too dim, even at a closer distance of 13.2 Mpc, the lower-limit distance in \citet{Kawabata2018PASJ...70..111K}. 

As previously discussed in \autoref{sec: spec_inputs}, SN\,2014dt was discovered past peak brightness and \citet{Kawabata2018PASJ...70..111K} estimated the bolometric light curve rise time in the range 11.8--21.0 days using other SN Iax. This range of rise times for SN\,2014dt is slightly longer than what is expected for either n1def or n3def, though it is largely consistent with other deflagration models in \citet{Fink2014MNRAS}. The decline rate $\Delta m^{B}_{15}$ estimated by \citet{Kawabata2018PASJ...70..111K} is 1.43--1.57 mag, while  \citet{Singh2018MNRAS.474.2551S} obtain 1.35 $\pm$ 0.06 mag; both of these decline rates are slower than the n1def and n3def models predict. Overall, the light-curve properties of SN\,2014dt suggest a slightly brighter and more slowly evolving object than the n1def model, and to a lesser extent the n3def model. Thus, there is a slight discrepancy between the results of our spectral modeling (favoring n1def) and the photometric properties of SN\,2014dt; we encourage further research into whether updated deflagration models can reconcile this issue.

The total ejecta masses, $M_{\rm ejecta}$, predicted for n1def and n3def are 0.0843 M$_{\odot}$ and 0.195 M$_{\odot}$, respectively.  This is within the 0.08--0.35 M$_{\odot}$ range found by \citet{Kawabata2018PASJ...70..111K} but far below the 0.95 M$_{\odot}$ predicted by \citet{Singh2018MNRAS.474.2551S}. This difference is due to the different assumptions and methods used in each paper, but is primarily driven by the higher distance and thus higher luminosity adopted by \citet{Singh2018MNRAS.474.2551S}. Both papers use a similar analysis of the late-time bolometric light curve to calculate the amount of $^{56}$Ni ($M_{\rm Ni56}$) synthesized in SN\,2014dt, but come up with different results. \citet{Singh2018MNRAS.474.2551S} suggest at least 0.14 M$_{\odot}$ of $^{56}$Ni was synthesized while \citet{Kawabata2018PASJ...70..111K} estimate a smaller M$_{\rm Ni56}$ of 0.035--0.105 M$_{\odot}$, which is a close match to both n1def and n3def. The important explosion parameters, $M_{\rm ejecta}$ and $M_{\rm Ni56}$, calculated by \citet{Kawabata2018PASJ...70..111K} for SN\,2014dt suggest it is consistent with both the n1def and n3def models while the higher $M_{\rm ejecta}$ and $M_{\rm Ni56}$ from \citet{Singh2018MNRAS.474.2551S} are more consistent with the higher densities of the n20def model, though we find that this model is not a good fit to the early-time spectra (see discussion in \autoref{sec: spec_results_analysis}). This again leads us to favor a lower distance and luminosity for SN\,2014dt.

We explored fitting SN\,2014dt with deflagration models having more ignition points (n10def, n20def, n150def, n1600def) that span the full breadth of model densities and abundances from \citet{Fink2014MNRAS}. The flattening of abundances in velocity space is a reasonable simplification for models up to and including n10def; n20def and higher models were left with their velocity-dependent abundances (note that the adjustments of removing elements past Zn, and the increase of $^{23}$Na, still remain). This means that the densities of models with low ignition points (n10def and lower) were able to be extrapolated down to zero velocity with a cubic spline, while n20def and above were not extrapolated. This limits our modelling of n20def and higher models to $\sim$ +420 days, instead of more than +500 days, since that is when \vphot\ dips below the velocity limits of those models. By this very late epoch, all deflagration models look remarkably similar to one another and we do not expect n20def and higher models to deviate much from lower models past $\sim$ +420 days. All other \tardis\ input parameters were the same as n1def for each epoch.

As before, the biggest differences between these deflagration models with more ignition points is seen at early times. At the earliest epoch, +11 days, n10def and higher models resemble n3def with an overall slightly worse fit. N20def and higher models were extremely similar, even at our earliest epoch of +11 days, with a small but noticeable difference in the strength of absorption at $\sim 5500$ \AA\ (n20def has the strongest absorption with decreasing strength toward n1600def). We find that past +64 days, all deflagration models produce a very similar spectrum -- with n1def and n3def fitting the blue end of the spectrum slightly better than higher models -- and are all able to do a reasonable job at fitting the optical spectrum of SN\,2014dt. This reinforces our earlier conclusion that the choice of densities in the range $\sim 1000$--7500 km s$^{-1}$ is a key input parameter for modelling the spectra of SN\,2014dt. Though our models do not seem to be as sensitive to the density profile below $\sim 1000$ km s$^{-1}$, this could be partly due to the overall very low densities for epochs past +64 days. 

\subsection{Fitting Strong Resonant Lines}\label{sec: resonant_lines}

We increase $^{23}$Na from the original n1def abundance of $\sim 0.01$\% to $\sim 1.7$\% in order to fit the strong Na I $\lambda\lambda$5890, 5896 doublet absorption in SN\,2014dt. A smaller $^{23}$Na abundance of $\sim 1\%$ was sufficient for reproducing the absorption feature for very late epochs, $>$ +400 days, but insufficient for reproducing the feature at +60--275 days. A $^{23}$Na abundance of $\sim 1.7$\% is high, but reasonable because the model isotopes reflect only the nucleosynthetic yield of a pure carbon-oxygen WD and do not include any preexisting $^{23}$Na in the system. We can speculate that a lower $^{23}$Na abundance is possible if its distribution is not spherically symmetric or uniform across all ejecta layers. In particular, the inner ejecta layers that correspond to the epochs that require more $^{23}$Na (300--2000 km s$^{-1}$) may have a higher $^{23}$Na abundance than outer layers. This stratified abundance structure is possible to model with \tardis\ but is beyond the scope of this work. 

Using the $v_{\rm outer}$ from the \citet{Fink2014MNRAS} models produced some spectral features that were too broad and blueshifted compared to the observed spectrum. Our choice of $v_{\rm outer}$, always less than the \citet{Fink2014MNRAS} model $v_{\rm outer}$, is based on fitting the feature that was most sensitive to the choice of $v_{\rm outer}$: the Ca II IR-triplet. In fact, this was the only noticeable feature affected after +200 days. In earlier epochs, Fe II features around 5000~\AA\ were also slightly weak and blueshifted. This strong Ca sensitivity to choice of $v_{\rm outer}$ may point to Ca being stratified -- requiring less Ca in the outer layers -- or not being spherically symmetric. Another possibility is that the Ca II IR-triplet is sensitive to the line treatment and might require non-LTE effects, which is not included in our radiative transfer model, as is suggested by \citet{Kasen_2006}. Further analysis of the effects these options might have on the Ca distribution, especially beyond $v_{\rm outer}$, could be the subject of future work. Our lower $v_{\rm outer}$ values also support the possibility that SN Iax may need steeper density profiles, relative to pure deflagration models, as has been suggested by \citet{Barna2018_mnras} and \citet{magee2022MNRAS.509.3580M}.

\subsection{Evolution of the Photosphere: Signs of a Quasi-Steady-State Wind}\label{sec: v_phot results}

\begin{figure*} 
\centering
\includegraphics[width=\linewidth]{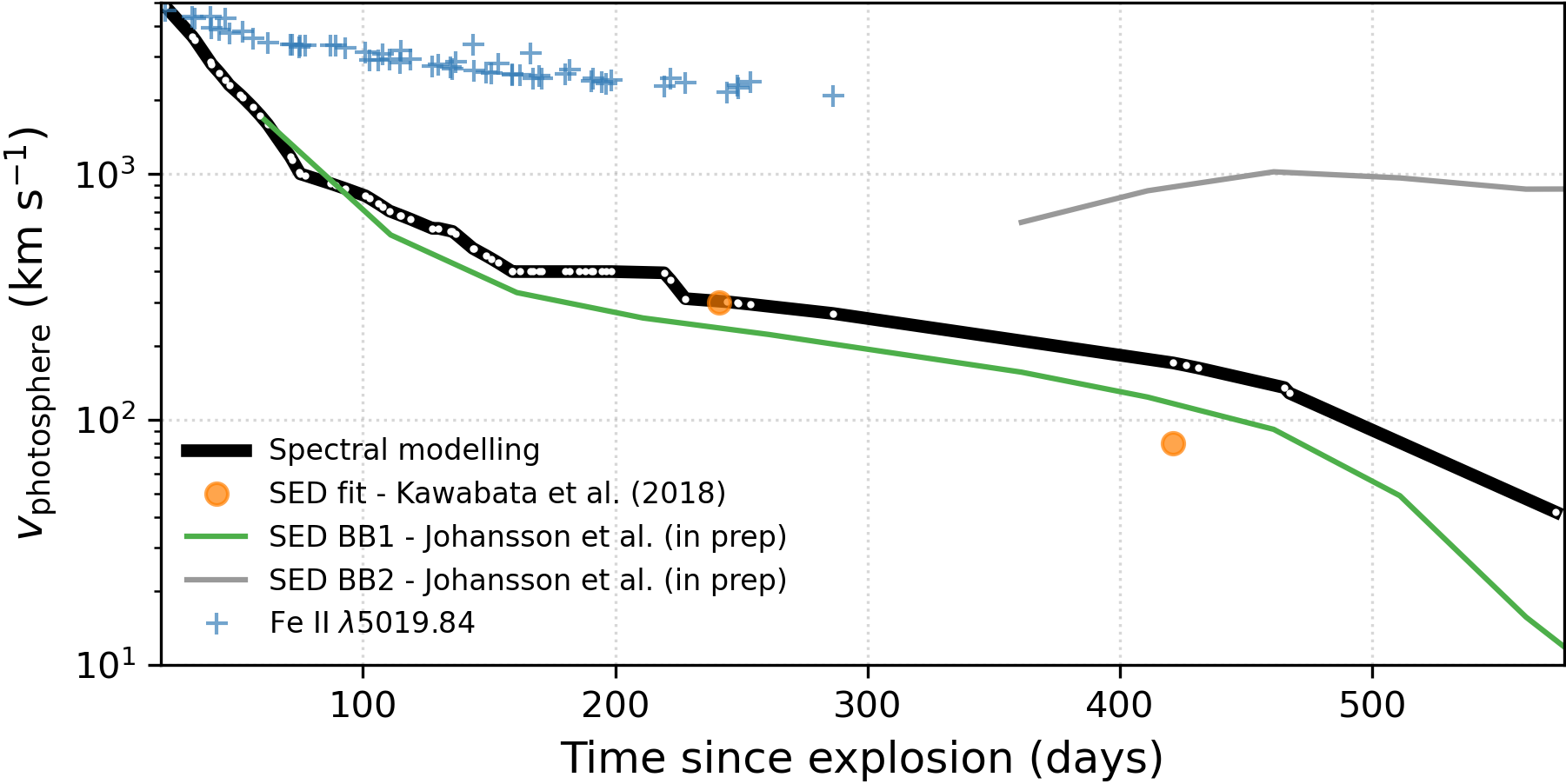}
\caption{The photospheric velocity, \vphot, of SN\,2014dt from modelling the optical spectra is shown as a solid black curve, with individual spectral fits as white dots. The line velocities of Fe II $\lambda$5019.84 are shown as blue crosses. For comparison, we display results from \citet{Kawabata2018PASJ...70..111K} (solid orange dots) who calculated the \vphot\ of SN\,2014dt using the bolometric luminosity and temperature from an analysis of the optical and near-IR photometry. We also show results from Johansson et al. (in prep.); they calculate \vphot\ from SED blackbody fits to the optical and near-IR photometry of SN\,2014dt. Their blackbody curve 1 (BB1; solid green) fits the optical and near-IR SED below $\sim 2$~$\mu$m while their blackbody curve 2 (BB2; solid gray) fits the SED above $\sim 2$~$\mu$m.}
\label{fig: v_phot}
\end{figure*}

With our spectral modelling we are able to obtain a robust evolution of the photospheric velocity, \vphot, that takes into account both the underlying thermal continuum as well as the complexity of fitting individual spectral lines. Our \vphot\ results from our best-fit model are listed in \autoref{tab: model_parameters} and plotted in \autoref{fig: v_phot}. 

For SN\,2014dt, we find that the evolution of \vphot\ is not simply fit by a pure exponential, power law, or linear function; though the fast decline of \vphot\ up to +150--200 days can be roughly fit by an exponential or power law, the subsequent slower evolution is better fit linearly in time. The \vphot\ of SN\,2014dt decreases very rapidly from 4800 km s$^{-1}$ to $\sim 1000$ km s$^{-1}$ from +11 to +64 days, and continues to decrease more slowly until it is $\sim 400$ km s$^{-1}$ by +148 days. \vphot\ then plateaus at $\sim 400$ km s$^{-1}$ for the next 50 days (we fit 16 spectra during this time) before slowly decreasing to $\sim 130$ km s$^{-1}$ by +456 days. Since we have two spectra (+454 and +456 days) from two different telescopes at roughly the same epoch, we are confident in both our reduction of the observed spectra and our model's ability to fit the spectra at this epoch. For +562 days, we obtain a \vphot\ of $\sim 40$ km s$^{-1}$. While we were able to model the continuum and some photospheric features in this last spectrum, the fit was overall worse; in particular, the strength of the Ca IR-triplet was higher in our model than what was observed. Owing to the faintness of SN\,2014dt by this time, and without any other spectrum at a similar phase, it may be that this last spectrum has some contamination from its host galaxy which could alter the continuum and the \vphot\ of our best-fit spectral model by tens of km s$^{-1}$. We also note that the low velocities we find in our late-time models are now more sensitive to any peculiar motion of the SN relative to its host, which we do not account for. 

\begin{figure} 
\centering
\includegraphics[width=\linewidth]{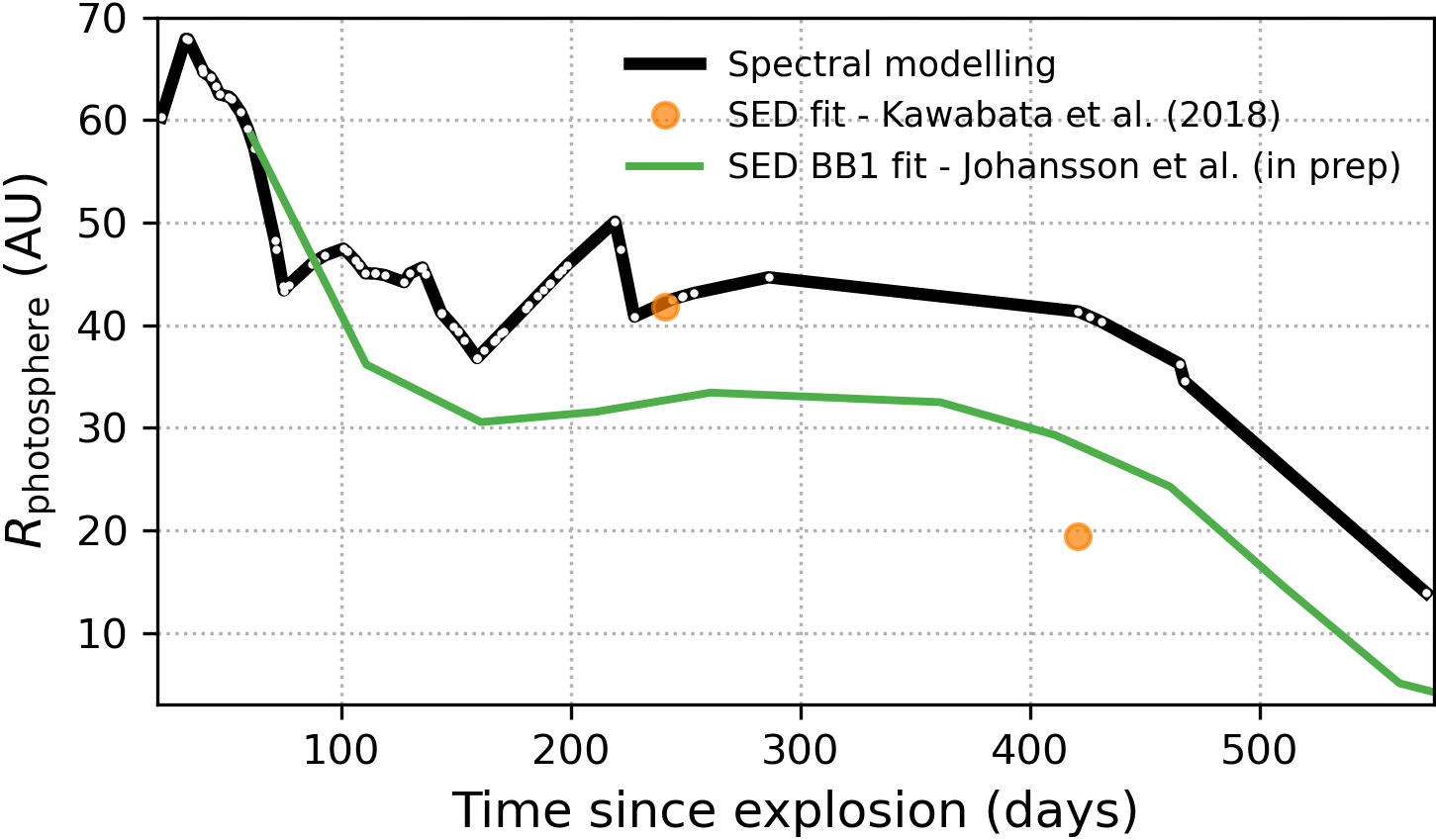}
\caption{Photospheric radius, $R_{\rm phot}$, of SN Iax SN\,2014dt from spectral models. For comparison, results from the photometric analysis by \citet{Kawabata2018PASJ...70..111K} and Johansson et al. (in prep.) are included.}
\label{fig: r_phot}
\end{figure}

With the assumption of homologous expansion, we also show the evolution of the photospheric radius $R_{\rm phot}$ in \autoref{fig: r_phot}. In order to convert the original $R_{\rm phot}$ results from \citet{Kawabata2018PASJ...70..111K} and Johansson et al. (in prep.) to \vphot, we add our adopted rise time, 11 days, to the phase in order to get a time since explosion, $t_{\rm exp}$. Assuming homologous expansion, \vphot\ = $R_{\rm phot}$ $\times$ $t_{\rm exp}^{-1}$.

We compare our \vphot\ results from spectral fitting with other works that derive \vphot\ from the photometry of SN\,2014dt. Our results are in excellent agreement with \citet{Kawabata2018PASJ...70..111K} at +230 days; they derive the bolometric luminosity and temperature using optical and near-IR photometry, and find the photosphere at $R_{\rm phot} \approx 6.2  \times 10^{14}$ cm ($\sim 300$ km s$^{-1}$). Using the additional assumption that the bolometric luminosity follows the $^{56}$Co decay, they obtain $R_{\rm phot} \approx 2.8 \times 10^{14}$ cm ($\sim 80$ km s$^{-1}$) at +410 days. This second \vphot, $\sim 80$ km s$^{-1}$, is much lower than our \vphot\ of $\sim 170$ km s$^{-1}$ for that same epoch. Our results suggest $R_{\rm phot}$ is $\sim 6.2 \times 10^{14}$ cm at +410 days, essentially unchanged from +230 days. If our $R_{\rm phot}$ at +410 days is correct, this would suggest that either the bolometric luminosity of SN\,2014dt is holding steady between +230 and +410 days (assuming a constant temperature; \citealt{Kawabata2018PASJ...70..111K}), or that a modest decrease in temperature, $\sim 1000$\,K, is needed to match the decreased bolometric luminosity of SN\,2014dt at +410 days from \citet{Kawabata2018PASJ...70..111K}. We believe the second scenario to be more likely as the modest decrease in temperature would still produce a blackbody curve that closely matches the +410 day spectral energy distribution (SED) from \citet{Kawabata2018PASJ...70..111K}.

Our \vphot\ results also broadly agree with the photometric analysis of Johansson et al. (in prep.). They fit two blackbody components, BB1 and BB2, to the optical and near-IR SED of SN\,2014dt. BB1 is used mainly to fit the SED below $\sim 2$~$\mu$m, and BB2 fits the SED above $\sim 2$~$\mu$m after +350 days. Our \vphot\ results match well with the general evolution of the BB1 component, with our \vphot\ being slightly ($\lesssim 100$ km s$^{-1}$) lower than their \vphot\ after +100 days. The BB2 component is at a much larger radius than the photosphere derived from our spectral modeling. This component is associated with the MIR excess seen by \citet{2016ApJ.Fox...816L..13F}, and though our data and models cannot directly bear upon the question of whether this long-wavelength flux is from dust emission or optically thick wind interaction \citep{Foley2016}, it is clear that it involves faster-moving material at larger distances than the optical photosphere.

Traditionally, \vphot\ is measured by line shifts of individual line profiles. In \autoref{fig: SDEC_evolution_full}, we show how most line profiles are largely blended and, with further exploration of the individual line interactions, we note that labelling any absorption or emission feature with a single line identification is often misleading. We are able to identify one absorption feature that is largely from a single line for all epochs when it appears, Fe II $\lambda$5019.84, and measure its velocity with a Gaussian fit to the line profile. At our earliest epoch, +11 days, the velocity measured from Fe II $\lambda$5019.84 agrees well with our results from spectral modelling. In the next epoch, just 10 days later, the velocity of Fe II is already $\sim 700$ km s$^{-1}$ higher than our spectral modelling \vphot. By +275 days, the last epoch where we could reliably identify Fe~II $\lambda$5019.84, the line velocity was $\sim 2000$ km s$^{-1}$ while the \vphot\ from spectral modelling was $\sim 270$ km s$^{-1}$. Overall, the velocity measured from Fe~II $\lambda$5019.84 declines much more slowly than our spectral modelling \vphot, suggesting that the Fe~II line is tracing faster, outer material in the ejecta which is distinct from the true underlying photosphere. Our results also suggest that measuring \vphot\ from a single line is no longer appropriate a few weeks past maximum brightness. 

The line velocities of Fe II $\lambda$6149 and $\lambda$6247 in SN\,2014dt are measured by \citet{Kawabata2018PASJ...70..111K} from +21--120 days and are in rough agreement with our Fe~II $\lambda$5019.84 velocities. Our extension of the line velocity out to +275 days demonstrates that Fe II continues to stays above $\sim 2000$ km s$^{-1}$ even at these very late epochs. Our analysis of the Fe II $\lambda$6149 and $\lambda$6247 lines with our spectral models show that the Fe II $\lambda$6149 absorption line is correctly identified and generally uncontaminated, but the Fe II $\lambda$6247 absorption is actually mostly Fe II $\lambda$6240.12. This different identification would only cause a small change in the resulting velocities,  $\sim 300$ km s$^{-1}$, which is within the $\sim 400$ km s$^{-1}$ line velocity scatter noted by \citet{Kawabata2018PASJ...70..111K}. Though we are able to confirm the line velocities from \citet{Kawabata2018PASJ...70..111K}, we find that the Fe II $\lambda$6149 and $\lambda$6247 lines have weaker absorption strengths and last for less time than Fe II $\lambda$5019.84. We identify Fe II $\lambda$5019.8, a seldom used, but very useful line that can provide an approximate \vphot\ value for very early times around maximum light and can also trace a physically important region for Fe out to later times than more commonly used lines can. 

 Based on \autoref{fig: r_phot}, we see that $R_{\rm phot}$ holds very steady at $\sim 40$--50 AU from +60 to 420 days and then decreases to $\sim 14$ AU by +562 days. While it is hard to be certain about the significantly lower $R_{\rm phot}$ at +562 days -- the \vphot\ uncertainty at this epoch was discussed at the beginning of this section -- the $R_{\rm phot}$ falling to $\sim 35$ AU by +454 days may indicate the smaller size at this epoch is real. These results are in general agreement with the evolution of the BB1 component from Johansson et al. (in prep.), though our $R_{\rm phot}$ estimate is $\sim 10$ AU higher after $\sim$ +100 days. Our results also concur with the estimate from \citet{Kawabata2018PASJ...70..111K} at +230 days; at +410 days, our $R_{\rm phot}$ is a factor of 2 larger than their estimate, but we discuss how a reasonable decrease in temperature could bring their estimate in agreement with ours earlier in this section. We do not include the BB2 component from Johansson et al. (in prep.) and the Fe II $\lambda$5019.84 line in \autoref{fig: r_phot} because they are poor tracers of the optical photosphere and produce an $R_{\rm phot}$ that is significantly larger than our spectral modelling results.

A photosphere with a constant velocity and density could be evidence for a steady-state wind coming from a bound remnant. Our results support this interpretation since we see a constant \vphot\ from +148 to 208 days and then a very slow decrease from +208 to 456 days. The wind model is further supported by the constant density seen at $\sim$ +150--450 days inferred from the steady ratio of forbidden to permitted Ca~II in the spectra of SN\,2014dt (other SN Iax also show a steady late-time density; \citealt{McCully2014ApJb}). A constant density breaks the \tardis\ model assumption of simple homologous expansion where density decreases as $t^{-3}$, and suggests more sophisticated models may be required. Even though \tardis\ cannot directly model the steady-state wind, we can estimate a wind mass-loss rate starting at $\sim$ +150 days when the photospheric velocity levels off. At this epoch \vphot\ = 400 km s$^{-1}$, $R_{\rm phot} = 5.5 \times 10^{14}$ cm, and $\rho$ at \vphot\ $= 1.7 \times 10^{-15}$ g cm$^{-3}$; these correspond to a wind mass-loss rate of $\sim 4 \times 10^{-3}$ M$_{\odot}$ yr$^{-1}$. If this mass-loss rate remains steady out to $\sim$ +450 days, the total mass in this wind would be $\sim 3 \times 10^{-3}$ M$_{\odot}$, a small fraction of the mass of the bound remnants found in weak deflagration models \citep{Fink2014MNRAS}. The slow decrease in \vphot\ over the range +208--456 days implies that the total mass in the wind may be a bit lower than this estimate. The drop in photospheric velocity at late times further suggests that the power source of the wind might be waning, perhaps consistent with the depletion of radioactive material. 

\section{Summary}\label{sec:summary}

We present an extensive spectral timeseries -- 69 total (21 new) spectra -- of the Type Iax SN\,2014dt, making it one of the best spectroscopically observed SN. The spectra show over 500 days of evolution in the long-lasting photospheric phase of SN Iax. In no other class of SN do photospheric lines persist to such late times. We identify the epochs between +90 and +150 days past maximum brightness as the spectral divergence phase, when SN Iax transition away from typical SN Ia spectral evolution.  

We are able to reproduce the optical spectra of SN\,2014dt with a nearly unaltered, weak deflagration explosion model (n1def) from \citet{Fink2014MNRAS} using the radiative transfer code \tardis. Our models are self-consistent, using the same chemical abundance and density profiles throughout, and are able to model the spectra from +11 to +562 days after $B_{\rm max}$. To our knowledge, this is the latest-epoch SN spectrum \tardis\ has ever been able to successfully model, only possible because of the continuing photospheric features of SN Iax at epochs where normal SN Ia are dominated by nebular features. It is astonishing that the n1def deflagration model, created well before the appearance of SN\,2014dt, can successfully reproduce the spectral evolution of this SN over 500 days in its evolution; such a match between a supernova model prediction and observations may be unprecedented.

We explore in detail the evolution of the photospheric velocity -- our main fitting parameter, \vphot. We find that the \vphot\ of SN\,2014dt decreases very rapidly from +11 to +64 days, then slows down significantly between +64 and +148 days, and continues to have a very slow, sometimes nonchanging, decrease until +456 days. The significant slowdown in the evolution of \vphot\ between +64 and +148 days closely overlaps the spectral divergence phase we find for SN\,2014dt, +90 to +150 days. The transition happens at a photospheric velocity of $\sim 400$--1000 km s$^{-1}$, which at these epochs corresponds to a photospheric radius of  $\sim 40$--50 AU. The correspondence in time between this slowdown and the spectral divergence phase suggests that this range of photospheric velocities may demarcate a boundary in the ejecta below which SN~Iax most strongly differ from their normal SN~Ia cousins.

The slow, nearly flat, evolution of \vphot\ from our spectral models, combined with a roughly constant density implied by the steady ratio of forbidden to permitted Ca~II, is consistent with the picture that a leftover bound remnant from SN\,2014dt began to drive an optically thick, quasi-steady-state wind. We estimate a wind mass-loss rate of a few times $10^{-3}$ M$_{\odot}$ yr$^{-1}$, corresponding to a total mass loss of a few times $10^{-3}$ M$_\odot$, during this phase from approximately +150 to +450 days after maximum light. Intriguingly, after about +450 days, the photospheric velocity begins to decrease more rapidly, leading us to speculate that the wind may be weakening, perhaps as the radioactive power source decays away. Future observations of more SN~Iax will illuminate this issue. Of special interest is whether the IR excess emission seen in SN\,2014dt \citep[][Johansson et al., in prep.;]{2016ApJ.Fox...816L..13F}, starting at about the same time as the wind dissipates (but seemingly originating at a much larger photospheric radius), will be commonly seen in SN~Iax. \emph{JWST} IR observations of SN~Iax may play an important role in answering these questions. 

\section*{Acknowledgements}

We thank Joe Lyman for providing the host-galaxy subtracted VLT/MUSE spectrum of SN\,2014dt, and Rob Fesen for the SALT/RSS spectrum of SN\,2014dt on 2015-04-13 (Dartmouth College program 2014-2-SCI-034). 

Spectra presented in this paper were obtained via the following programs: Rutgers University SALT programs 2014-1-MLT-001 and 2015-1-MLT-002 (PI S. Jha); UC Berkeley Keck programs U048D and U078D (PI A. Filippenko); NASA Keck programs 2014B/N125D and 2015A/N129D (PI S. Jha); LCO Supernova Key Project KEY2014A-003 (PI A. Howell); SOAR program N2015A-0253 (PI R. Foley); and Gemini program GN-2015B-Q-78 (PI S. Jha).

This research made use of \tardis, a community-developed software package for spectral
synthesis in supernovae \citep{Kerzendorf2014MNRAS.440..387K, kerzendorf_wolfgang_2023_7525913}. The
development of \tardis\ received support from GitHub, the Google Summer of Code
initiative, and from ESA's Summer of Code in Space program. \tardis\ is a fiscally
sponsored project of NumFOCUS. \tardis\ makes extensive use of Astropy and Pyne.
This research has made use of the \citet{https://doi.org/10.26132/ned1}, which is funded by NASA and operated by the California Institute of Technology.
This work made use of the Heidelberg Supernova Model Archive (HESMA), https://hesma.h-its.org

This research has made use of the Keck Observatory Archive (KOA), which is operated by the W. M. Keck Observatory and the NASA Exoplanet Science Institute (NExScI), under contract with the National Aeronautics and Space Administration (NASA). The Keck Observatory is operated as a scientific partnership among the California Institute of Technology, the University of California, and NASA; it was made possible by the generous financial support of the W. M. Keck Foundation.
This research uses services or data provided by the Astro Data Archive at NSF's NOIRLab. NOIRLab is operated by the Association of Universities for Research in Astronomy (AURA), Inc. under a cooperative agreement with the National Science Foundation (NSF).
This work uses data from the Las Cumbres Observatory (LCO) Global Telescope Network. The LCO group is supported by NSF grant AST-1911225.
A major upgrade of the Kast spectrograph on the Shane 3~m telescope at Lick Observatory, led by Brad Holden, was made possible through generous gifts from the Heising-Simons Foundation, William and Marina Kast, and the University of California Observatories. 
Research at Lick Observatory is partially supported by a generous gift from Google.
We appreciate the excellent assistance of the staffs of the various observatories at which data were obtained.

Y.C.-N. and S.W.J. are grateful for support from NSF awards AST-0847157 and AST-1615455, and NASA grant HST-GO-16683. L.A.K. acknowledges support by NASA FINESST fellowship 80NSSC22K1599. C.L. acknowledges support from the NSF Graduate Research Fellowship under grant DGE-2233066.
B.B. is supported by the \'UNKP-22-4 - SZTE-476 New National Excellence Program of the Ministry for Culture and Innovation from the source of the National Research, Development and Innovation Fund.
T.S. is supported by the NKFIH/OTKA FK-134432 grant of the National Research, Development and Innovation (NRDI) Office of Hungary, the J\'anos Bolyai Research Scholarship of the Hungarian Academy of Sciences and by the New National Excellence Program (\'UNKP-22-5) of the Ministry for Innovation and Technology of Hungary from the source of NRDI Fund.
A.V.F. acknowledges financial assistance from the Christopher R. Redlich Fund and numerous individual donors.

\software{\texttt{TARDIS} \citep{Kerzendorf2014MNRAS.440..387K, kerzendorf_wolfgang_2023_7525913}, Jupyter \citep{Jupyter_soton403913}, Astropy \citep{AstropyCollaboration2013, AstropyCollaboration2018}, \texttt{IRAF} \citep{IRAF1, IRAF2}, Matplotlib \citep{Hunter2007}, NumPy \citep{harris2020array}, pandas \citep{McKinney2010}, SciPy \citep{2020SciPy}, Pyraf \citep{Pyraf}}

\begin{longtable*}{ccccc|c}

\caption{\texttt{TARDIS} parameters} \label{tab: model_parameters} \\

\hline \hline 
Phase$^{a}$ & $t_{\rm exp }$ & $L_{\rm{requested}}$ & $L$: $\lambda_{\rm{min}}$ -- $\lambda_{\rm{max}}$ & $v_{\rm outer}$ & $v_{\rm phot}$\\ 
 (days) & (days) & (L$_{\odot}$) & (\AA) & (km s$^{-1}$) & (km s$^{-1}$)\\ \hline 
\endfirsthead

\multicolumn{6}{c}%
{{\bfseries \tablename\ \thetable{} -- (continued)}} \\
\hline \hline 
Phase$^{a}$ & $t_{\rm exp }$ & $L_{\rm{requested}}$ & $L$: $\lambda_{\rm{min}}$ -- $\lambda_{\rm{max}}$ & $v_{\rm outer}$ & $v_{\rm phot}$\\ 
 (days) & (days) & (L$_{\odot}$) & (\AA) & (km s$^{-1}$) & (km s$^{-1}$)\\ \hline 
\endhead

\hline 
\multicolumn{6}{r}{{Continued on next page}} \\ 
\endfoot

\hline
\multicolumn{6}{l}{{$^{a}$ Days since B maximum, 2014 October 20.4 (MJD 56950.4),}}\\
\multicolumn{6}{l}{{corrected for time dilation }}\\

\endlastfoot
+11 & 21.7 & \num{1.65e+08} & 4064 -- 8757 & 7502 & 4800\\
+21 & 32.3 & \num{1.35e+08} & 4479 -- 8951 & 7507 & 3637\\
+22 & 33.5 & \num{1.32e+08} & 4500 -- 8452 & 7507 & 3513\\
+28 & 39.5 & \num{1.05e+08} & 4280 -- 8452 & 7510 & 2855\\
+29 & 40.0 & \num{1.27e+08} & 3982 -- 9446 & 7510 & 2800\\
+32 & 43.1 & \num{4.80e+07} & 5328 -- 6970 & 7511 & 2580\\
+34 & 45.4 & \num{9.11e+07} & 4280 -- 8452 & 7512 & 2417\\
+36 & 47.0 & \num{1.03e+08} & 3982 -- 9446 & 7513 & 2304\\
+40 & 51.3 & \num{4.96e+07} & 5808 -- 8951 & 7284 & 2103\\
+41 & 52.4 & \num{7.37e+07} & 4300 -- 8452 & 7225 & 2051\\
+45 & 56.3 & \num{6.37e+07} & 4363 -- 8452 & 7016 & 1869\\
+48 & 59.4 & \num{2.22e+07} & 6469 -- 8453 & 6852 & 1727\\
+51 & 62.3 & \num{2.33e+07} & 4500 -- 6000 & 6696 & 1591\\
+60 & 71.2 & \num{4.36e+07} & 4300 -- 8452 & 6218 & 1175\\
+61 & 71.9 & \num{3.54e+07} & 5098 -- 9296 & 6179 & 1140\\
+64 & 74.8 & \num{4.58e+07} & 3982 -- 9446 & 4901 & 1015\\
+64 & 75.1 & \num{3.46e+07} & 5009 -- 8951 & 4795 & 1000\\
+66 & 77.3 & \num{1.46e+07} & 4145 -- 5965 & 4782 & 984\\
+76 & 87.2 & \num{2.84e+07} & 4465 -- 8452 & 4721 & 913\\
+78 & 89.0 & \num{3.17e+07} & 4280 -- 8452 & 4710 & 900\\
+82 & 93.0 & \num{2.13e+07} & 5032 -- 8951 & 4685 & 871\\
+90 & 101.1 & \num{2.60e+07} & 4170 -- 8452 & 4636 & 813\\
+92 & 102.6 & \num{2.70e+07} & 4181 -- 9446 & 4626 & 795\\
+95 & 106.2 & \num{2.84e+07} & 3973 -- 8861 & 4605 & 755\\
+97 & 107.9 & \num{2.25e+07} & 4377 -- 8452 & 4594 & 735\\
+100 & 110.5 & \num{2.13e+07} & 4185 -- 9446 & 4578 & 706\\
+104 & 114.8 & \num{1.59e+07} & 4714 -- 8951 & 4552 & 679\\
+104 & 114.9 & \num{2.07e+07} & 4480 -- 8451 & 4551 & 678\\
+108 & 119.0 & \num{2.04e+07} & 4381 -- 8451 & 4526 & 653\\
+117 & 127.5 & \num{2.23e+07} & 3982 -- 9446 & 4478 & 600\\
+119 & 130.0 & \num{1.50e+07} & 4246 -- 8837 & 4480 & 600\\
+124 & 134.5 & \num{1.76e+07} & 4016 -- 9446 & 4483 & 586\\
+124 & 135.5 & \num{1.61e+07} & 4280 -- 9448 & 4484 & 583\\
+126 & 136.8 & \num{1.51e+07} & 4386 -- 8452 & 4485 & 569\\
+133 & 143.6 & \num{1.13e+07} & 4814 -- 8951 & 4490 & 497\\
+133 & 143.9 & \num{1.83e+07} & 3977 -- 8795 & 4490 & 495\\
+138 & 148.9 & \num{1.80e+07} & 3978 -- 8844 & 4494 & 464\\
+140 & 150.9 & \num{1.61e+07} & 3979 -- 8844 & 4496 & 452\\
+143 & 153.7 & \num{1.16e+07} & 4677 -- 8452 & 4498 & 435\\
+148 & 159.2 & \num{1.60e+07} & 3982 -- 9446 & 4502 & 400\\
+148 & 159.3 & \num{1.09e+07} & 5103 -- 9299 & 4502 & 400\\
+151 & 162.4 & \num{9.78e+06} & 4893 -- 8951 & 4502 & 400\\
+156 & 166.5 & \num{1.22e+07} & 4343 -- 8452 & 4502 & 400\\
+156 & 167.3 & \num{1.49e+07} & 3932 -- 9402 & 4502 & 400\\
+159 & 169.7 & \num{1.36e+07} & 3982 -- 8800 & 4502 & 400\\
+160 & 170.7 & \num{1.49e+07} & 3978 -- 8844 & 4502 & 400\\
+169 & 180.2 & \num{1.27e+07} & 3782 -- 9448 & 4502 & 400\\
+171 & 181.8 & \num{5.82e+06} & 4480 -- 6554 & 4502 & 400\\
+175 & 185.6 & \num{1.04e+07} & 3982 -- 8321 & 4502 & 400\\
+177 & 188.1 & \num{9.96e+06} & 4537 -- 9446 & 4502 & 400\\
+180 & 190.6 & \num{1.09e+07} & 4181 -- 8842 & 4502 & 400\\
+180 & 191.2 & \num{1.36e+07} & 3902 -- 9447 & 4502 & 400\\
+184 & 194.5 & \num{1.10e+07} & 4120 -- 8795 & 4502 & 400\\
+185 & 196.2 & \num{1.04e+07} & 3952 -- 9447 & 4502 & 400\\
+187 & 198.5 & \num{1.19e+07} & 3977 -- 8842 & 4502 & 400\\
+208 & 219.3 & \num{9.23e+06} & 3980 -- 8843 & 4502 & 395\\
+211 & 221.9 & \num{1.03e+07} & 3684 -- 9447 & 4502 & 370\\
+217 & 227.8 & \num{9.64e+06} & 3932 -- 9446 & 4502 & 310\\
+233 & 244.1 & \num{8.17e+06} & 3979 -- 8843 & 4501 & 302\\
+237 & 248.4 & \num{7.10e+06} & 4479 -- 8472 & 4501 & 298\\
+238 & 248.7 & \num{8.73e+06} & 4051 -- 9448 & 4501 & 298\\
+243 & 253.6 & \num{8.08e+06} & 5017 -- 9447 & 4501 & 294\\
+275 & 286.2 & \num{5.43e+06} & 4483 -- 8469 & 4501 & 270\\
+410 & 421.1 & \num{2.80e+06} & 5014 -- 9446 & 4500 & 170\\
+415 & 426.1 & \num{3.16e+06} & 3985 -- 9448 & 4500 & 166\\
+420 & 431.0 & \num{2.04e+06} & 5274 -- 9087 & 4500 & 162\\
+454 & 465.2 & \num{1.34e+06} & 5226 -- 8759 & 4500 & 135\\
+456 & 467.2 & \num{1.07e+06} & 5373 -- 8513 & 4500 & 128\\
+562 & 572.7 & \num{3.96e+05} & 4616 -- 9204 & 4499 & 42\\
\end{longtable*}

\bibliography{main}
\end{document}